%% file: ANA-HIGP-2024-32-PAPER.tex
\documentclass[PAPER, atlasdraft=false, USenglish, cernpreprint, texmf, orcidlogo]{atlasdoc}
\usepackage{atlaspackage}
\usepackage{comment}
\usepackage{multirow}
\usepackage{makecell}
\usepackage{atlasbiblatex}

\usepackage{atlasphysics}
\graphicspath{{logos/}{figures/}}

\usepackage{ANA-HIGP-2024-32-PAPER-defs}

\AtlasTitle{A search for triple Higgs boson production in the $6b$ final state using $pp$ collisions at $\sqrt{s}=13$ TeV with the ATLAS detector}

\AtlasAbstract{%
A search for the production of three Higgs bosons ($HHH$) in the $b\bar{b}b\bar{b}b\bar{b}$ final state is presented.
The search uses 126~$\text{fb}^{-1}$ of proton--proton collision data at $\sqrt{s}=13$ TeV collected with the ATLAS detector at the Large Hadron Collider.
The analysis targets both non-resonant and resonant production of $HHH$.
The resonant interpretations primarily consider a cascade decay topology of $X\rightarrow SH\rightarrow HHH$ with masses of the new scalars $X$ and $S$ up to 1.5 TeV and 1 TeV, respectively.
In addition to scenarios where $S$ is off-shell, the non-resonant interpretation includes a search for standard model (SM) $HHH$ production, with limits on the tri-linear and quartic Higgs self-coupling set.
No evidence for $HHH$ production is observed.
An upper limit of 59 fb is set, at 95\% confidence level, on the cross-section for Standard-Model $HHH$ production.
}

\AtlasRefCode{HIGP-2024-32}

\PreprintIdNumber{CERN-EP-2024-285}

\AtlasDate{\today}

\AtlasJournal{Phys.\ Rev.\ D.}
\AtlasJournalRef{\PRD 111 (2025) 032006}
\AtlasDOI{10.1103/PhysRevD.111.032006}

\AtlasCoverEgroupAnalysisTeam{atlas-hdbs-2022-07-analysis-team@cern.ch, atlas-higp-2024-32-analysis-team@cern.ch}

\hypersetup{pdftitle={ATLAS document},pdfauthor={The ATLAS Collaboration}}

\begin{document}

\maketitle

\section{Introduction}
\label{sec:intro}

%
Since the discovery of the \(\SI{125}{\GeV}\) Higgs boson (\(H\))~\cite{HIGG-2012-27,CMS-HIG-12-028} at the Large Hadron Collider's (LHC) ATLAS and CMS experiments, a core part of the LHC research program has been to investigate the properties of the Higgs boson and establish if they are in agreement with the predictions of the Standard Model (SM) \cite{HIGG-2021-23,CMS-HIG-22-001}.
Two key parts of the SM Higgs mechanism are the tri-linear (\(\lambda_{3}\)) and quartic (\(\lambda_{4}\)) self-coupling constants, which are a crucial part of the electroweak symmetry breaking mechanism and determine the shape of the Higgs field potential.
Higgs self-coupling modifiers \kthree\ and \kfour\ are introduced using the \(\kappa\) framework, in which they are defined as \(\kappa_i = \lambda_i / \lambda_i^{\text{SM}}\) for \(i=3,4\), such that \(\kappa_i=1\) corresponds to the couplings predicted by the SM \cite{LHC_HWG_recommendations}.
Combinations of searches for di-Higgs production constrain \kthree\ (also often referred to as \(\kappa_{\lambda}\)) to be between \(-1.2\) and 7.2 at 95\% confidence level (CL)~\cite{diHiggsPublished,HIGG-2021-23,CMS-HIG-22-001}.
The most recent combination of results for single- and di-Higgs production further constrains \kthree\ to be between \(-0.4\) and 6.3 (when other Higgs boson couplings are assumed to have their SM values)~\cite{HDBS-2022-03}.
However, there have been no experimental constraints on \(\lambda_{4}\) or its relation to \(\lambda_{3}\) prior to the results presented in this paper.
Searches for \(HHH\) production have been proposed in Refs. \cite{Papaefstathiou_2019,Papaefstathiou_2024}, which specifically identify the six \(b\)-jet final state as a suitable channel and explore the possibility of constraints on \(\lambda_{4}\) and \(\lambda_{3}\).
A recent comprehensive whitepaper \cite{hhhwhitepaper} explores additional channels and identifies key challenges for future \HHH~searches searches.

This paper presents searches for the simultaneous production of three Higgs bosons --- tri-Higgs (\HHH) --- in the six \(b\)-jet final state.
Data collected during Run 2 of the LHC by the ATLAS experiment~\cite{PERF-2007-01} between 2016 and 2018 are used.
As well as a unique dependence on \(\lambda_{4}\) and a dependence on \(\lambda_{3}\), \HHH\ provides new sensitivity to Beyond SM (BSM) physics with extended scalar sectors.
The results are interpreted in terms of three benchmark models.

\begin{figure}
\centering
\subfloat[]{\label{fig:intro:feynman:resonant_cascade}\includegraphics[width=0.3\textwidth]{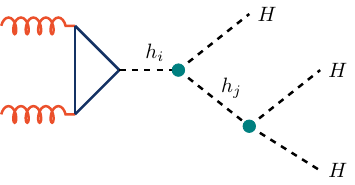}}\qquad
\subfloat[]{\label{fig:intro:feynman:triangle_quartic}\includegraphics[width=0.3\textwidth]{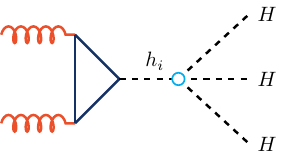}}\\
\subfloat[]{\label{fig:intro:feynman:square}\includegraphics[width=0.3\textwidth]{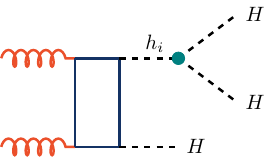}}\qquad
\subfloat[]{\label{fig:intro:feynman:pentagon}\includegraphics[width=0.3\textwidth]{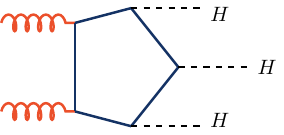}}
\caption{Leading-order Feynman diagrams for gluon--gluon fusion triple Higgs-boson production. The symbols \(h_i\) and \(h_j\) represent any of the SM Higgs \(H\) or heavy scalars, \(X\) and \(S\). In particular, the resonant cascade decay is shown in (a). For the case where \(h_i=h_j=H\) the solid circle indicates the tri-linear self-coupling \(\lambda_3\), and the open circle indicates the quartic-linear self-coupling \(\lambda_4\).}
\label{fig:intro:feynman}
\end{figure}

First, SM-like scenarios are considered to place constraints on the Higgs self-coupling modifiers \kthree\ and \kfour.
All leading-order gluon--gluon fusion (ggF) production modes represented in Figure~\ref{fig:intro:feynman} are included.

Second, scenarios extending the SM by adding two real scalar bosons \(X\) and \(S\) are probed, which are hypothesised in several BSM models.
The mass ranges \(325<\mX<\SI{575}{\GeV}\) and \(200<\mS<\SI{350}{\GeV}\) are considered and by convention \(\mX > \mS\).
In this interptretation the mass ranges are chosen to fulfill perturbative unitarity bounds~\cite{TRSM2021}.
The scenarios include cases in which either resonant or non-resonant production is dominant.
All leading-order ggF production modes (illustrated in \Fig{\ref{fig:intro:feynman}}) are considered, including interference between all diagrams, both BSM and SM.
Models including the Two Real Scalar Model (TRSM)~\cite{Robens_TRSM_2020,TRSM2021} and a ``simple model for dark matter and \(CP\) violation'' (DM-CPV) \cite{DMCPV2022} feature this extension to the SM.
For suitable parameter choices the \(HHH\) production cross-section is preferentially enhanced without increasing single- or di-Higgs production beyond experimental constraints.
Negligible differences are found in the \(HHH\) event kinematics between the TRSM and DM-CPV models, so Monte Carlo (MC) simulated events are generated only in the TRSM as described in \Sect{\ref{sec:dataMC}}, and results are interpretatable in both models.

Finally, scenarios with resonant production of generic heavy spin-0 bosons \(X\) and \(S\) are also considered, with masses \(\mX > \mS\), covering \(550<\mX<\SI{1500}{\GeV}\), and \(275<\mS<\SI{1000}{\GeV}\).
The search covers benchmark scenarios in which both particles have either narrow or wide decay width, with resonances that have a width below or just above the detector resolution, respectively.
Only resonant ggF production (Figure~\ref{fig:intro:feynman:resonant_cascade}) is chosen for consideration to give a simplified treatment and avoid the large contributions from non-resonant diagrams at leading order in the TRSM.
This interpretation allows the search to probe generically for resonances that lie beyond the masses permitted by the TRSM benchmark.

To enable these interpretations, the analysis is divided into three searches based on the kinematics of the signal models considered: ``Non-resonant'', ``Resonant'' and ``Heavy-resonant''.
The Non-resonant search is used for the TRSM parameter space dominated by non-resonant production and the SM-like interpretation, the Resonant search is used for the TRSM parameter space dominated by resonant production, and the Heavy-resonant search is used for the generic heavy resonance signal models.
Each search follows the same general analysis strategy.
Deep neural network (DNN) classifiers are trained to discriminate between the relevant signals and the SM background.
The dominant SM background process in all searches is Quantum Chromodynamics (QCD) multijet production, which is estimated by using an entirely data-driven approach based on an extrapolation between different \(b\)-jet multiplicities.
To obtain the final results in each search, a profile likelihood fit is performed over the binned DNN score distribution.

The structure of this paper is as follows.
In Section~\ref{sec:detector} the ATLAS detector is described, followed by a summary of the data and MC simulated events used in Section~\ref{sec:dataMC}.
Section~\ref{sec:objects} explains how events are reconstructed, and Section~\ref{sec:strategy} explains the analysis strategy and how signal events are selected.
In Section~\ref{sec:backgrounds}, the estimate of the SM background is described, followed by an explanation of the statistical interpretation in Section~\ref{sec:stats}.
A summary of the uncertainties considered is presented in Section~\ref{sec:systs} and finally the results and conclusions are provided in Sections~\ref{sec:results} and~\ref{sec:conclusion}, respectively.


%
\section{ATLAS detector}
\label{sec:detector}

%
\newcommand{\AtlasCoordFootnote}{%
ATLAS uses a right-handed coordinate system with its origin at the nominal interaction point (IP)
in the centre of the detector and the \(z\)-axis along the beam pipe.
The \(x\)-axis points from the IP to the centre of the LHC ring,
and the \(y\)-axis points upwards.
Polar coordinates \((r,\phi)\) are used in the transverse plane,
\(\phi\) being the azimuthal angle around the \(z\)-axis.
The pseudorapidity is defined in terms of the polar angle \(\theta\) as \(\eta = -\ln \tan(\theta/2)\) and is equal to the rapidity
$ y = \frac{1}{2} \ln \left( \frac{E + p_z}{E - p_z} \right) $ in the relativistic limit.
Angular distance is measured in units of \(\Delta R \equiv \sqrt{(\Delta y)^{2} + (\Delta\phi)^{2}}\).}
The ATLAS experiment~\cite{PERF-2007-01} at the LHC is a multipurpose particle detector
with a forward--backward symmetric cylindrical geometry and a near \(4\pi\) coverage in
solid angle.\footnote{\AtlasCoordFootnote}
It consists of an inner tracking detector (ID) surrounded by a thin superconducting solenoid
providing a \qty{2}{\tesla} axial magnetic field, electromagnetic and hadronic calorimeters, and a muon spectrometer.
The inner tracking detector covers the pseudorapidity range \(|\eta| < 2.5\).
It consists of silicon pixel, silicon microstrip, and transition radiation tracking detectors.
Lead/liquid-argon (LAr) sampling calorimeters provide electromagnetic (EM) energy measurements
with high granularity within the region \(|\eta|< 3.2\).
A steel/scintillator-tile hadronic calorimeter covers the central pseudorapidity range (\(|\eta| < 1.7\)).
The endcap and forward regions are instrumented with LAr calorimeters
for EM and hadronic energy measurements up to \(|\eta| = 4.9\).
The muon spectrometer (MS) surrounds the calorimeters and is based on
three large superconducting air-core toroidal magnets with eight coils each.
The field integral of the toroids ranges between \num{2.0} and \qty{6.0}{\tesla\metre}
across most of the detector.
The muon spectrometer includes a system of precision tracking chambers up to \(|\eta| = 2.7\) and fast detectors for triggering up to \(|\eta| = 2.4\).
The luminosity is measured mainly by the LUCID--2~\cite{LUCID2} detector, which is located close to the beampipe.
A two-level trigger system is used to select events~\cite{TRIG-2016-01}.
The first-level (L1) trigger is implemented in hardware and uses a subset of the detector information
to accept events at a rate below \qty{100}{\kHz}.
This is followed by a software-based trigger (HLT) that
reduces the accepted event rate to \qty{1}{\kHz} on average
depending on the data-taking conditions.
A software suite~\cite{SOFT-2022-02} is used in data simulation, in the reconstruction
and analysis of real and simulated data, in detector operations, and in the trigger and data acquisition
systems of the experiment.


%
\section{Data and Monte Carlo simulations}
\label{sec:dataMC}

%
\subsection{Data and triggers}
\label{sec:dataMC:data}

The proton--proton (\(pp\)) collisions used in this paper were collected between 2016 and 2018, at a center-of-mass energy of \(\sqrt s=13~\TeV\).
They correspond to an integrated luminosity of \(126~\ifb\), with an uncertainty of 0.84\%~\cite{DAPR-2021-01} obtained using the LUCID-2 detector for the primary luminosity measurements, complemented by measurements using the inner detector and calorimeters.
Only events that satisfy data quality requirements that ensure the stable operation of the ATLAS detector \cite{DAPR-2018-01} are considered.

Events from each year satisfy triggers requiring multiple jets and \(b\)-tagged jets~\cite{TRIG-2018-08}, which were implemented in 2016.
These jets are reconstructed using the \antikt\ algorithm with radius parameter \(R=0.4\)~\cite{Cacciari:2008gp,Fastjet}.
Jets originating from \(b\)-quarks are tagged using the \texttt{MV2c20} (\texttt{MV2c10}) algorithm in 2016 (2017--2018)~\cite{TRIG-2018-08} operating at \(60-70\%\) identification efficiency as measured using simulated \ttbar\ events.
For all years, the triggers used require four jets with transverse momentum \(\pT>\SI{15}{\GeV}\) and \(|\eta|<2.5\) at L1 level.
In 2016 and 2018, the trigger used requires two \(b\)-tagged jets and two additional jets at HLT level, all with \(\pT>\SI{35}{\GeV}\) (2b+2j).
In 2017, a trigger requiring three \(b\)-tagged jets and one additional jet at HLT level, all with \(\pT>\SI{35}{\GeV}\) (3b+1j) is used instead.
The triggers selected maximize the efficiency for the resonant (\mX, \mS) phase space within the TRSM perturbative unitarity bounds.
Different triggers were used in different years due to changes in the online \(b\)-tagging working points.
Depending on the mass point and data-taking year, the triggers used are 90--98\% efficient.
During 2016, a fraction of the data taking (\(8.3~\ifb\)) was affected by an inefficiency in the trigger-level vertex reconstruction, which reduced the efficiency of the algorithms used to identify \(b\)-tagged jets; those events are not retained for further analysis.

\subsection{Signal models and event simulation}
\label{sec:dataMC:MC}

For all signal samples described in this section, MC simulated signal events are generated using \MADGRAPH~\cite{Alwall:2011uj} 2.9.5 at leading-order in the strong coupling constant, using the \NNPDF{3.0NLO} parton distribution function (PDF) set~\cite{Ball:2014uwa} with the A14 set of tuned parameters (tune)~\cite{ATL-PHYS-PUB-2014-021}.
The \(H\rightarrow b\bar{b}\) branching fraction is taken to be 0.582, and its total width \(\SI{4.088}{\MeV}\) corresponding to the SM values for a Higgs boson mass of \(\SI{125}{\GeV}\)~\cite{deFlorian:2016spz}.
All events are showered with \PYTHIA\ 8.245~\cite{Sjostrand:2014zea}, and a generation-level filter requiring at least four \(b\)-tagged\footnote{At truth-level, \(b\)-tagged jets are defined as jets which have a \(b\)-hadron of \(\pT>\SI{5}{\GeV}\) lying within a cone of size \(\Delta R=0.3\) from their axis.} jets with \(\pt>\SI{25}{\GeV}\) is applied.

For the SM-like interpretation in the Non-resonant search a single sample is generated at \((\kthree, \kfour)=(1,1)\).
All diagrams in Figure~\ref{fig:intro:feynman} are included, fully accounting for interference between them.
The Higgs bosons are decayed exclusively to \(b\bar{b}\).
A reweighting \cite{NLOMadGraphReweighting} allows the sample kinematics to be altered to reflect different values of \kthree\ and \kfour, which follows a similar procedure to that outlined in Ref.~\cite{Papaefstathiou:2015paa}.
The SM \(HHH\) ggF production cross-section at a center-of-mass energy of \(\sqrt{s}=\SI{13}{\TeV}\) is taken as \(\sigma_{HHH}^{SM}=0.079^{+0.012}_{-0.013}\SI{}{fb}\) at next-to-next-to-leading-order (NNLO), calculated by extrapolating cross-sections at higher \(\sqrt{s}\) values presented in Ref.~\cite{HHH-NNLO-14TeV} and including uncertainties from missing higher-order corrections (\(+5\%, -8\%\)) and the finite top quark mass approximation (\(\pm15\%\)).

For the Resonant and Non-resonant searches, Benchmark 3 of the TRSM, defined in Ref.~\cite{TRSM2021}, is used for signal generation\footnote{A more recent benchmark satisfying additional constraints from the electroweak phase transition is presented in Ref.~\cite{Karkout:2024ojx}, but the kinematic differences between the new benchmark and the one used here are minimal.}.
Here, the SM is extended by two neutral CP-even scalars, \(X\) and \(S\).
Interactions between \(X/S\) and non-Higgs SM particles come via the mixing between the \(X/S\) fields and the SM Higgs field, making the couplings proportional to the SM-Higgs couplings.
Here, by choice, \(\mX\geq\mS\geq\mH\).
The benchmark chosen sets the TRSM parameters to maximize \(HHH\) production through a resonant cascade decay \(gg \rightarrow X \rightarrow HS \rightarrow HHH\), though all possible leading-order \(gg\rightarrow HHH\) diagrams are generated together with interference.
Typical values for the branching fractions \(X\rightarrow SH\) and \(S\rightarrow HH\) are 45\% and 70\%, respectively.
For more information see Ref~\cite{TRSM2021}.

For the TRSM signals, the Higgs bosons are decayed inclusively using \PYTHIA\ 8.245~\cite{Sjostrand:2014zea}.
A set of 27 \((\mX, \mS)\) points are chosen within the perturbative unitarity bounds of the TRSM, with \(325~\leq~\mX~\leq~\SI{575}{\GeV}\) and \(200~\leq~\mS~\leq~\SI{350}{\GeV}\).
This includes two categories: 19 ``resonant'' points where \(\mX>\mS+\mH\) and \(\mS>2\mH\) such that the resonant cascade decay is on-shell and dominant, and eight ``non-resonant'' points where \(\mS\leq 2\mH\) such that the resonant cascade is off-shell and non-resonant diagrams become dominant.

The results of the Resonant and Non-resonant searches can also be intepreted in the DM-CPV model, presented in Ref.~\cite{DMCPV2022}, using benchmarks 1 and 3 defined therein.
These benchmarks are defined to maximize \(HHH\) production, and can be extended across the same \((\mX,\mS)\) plane as the TRSM.
In the nomenclature of Ref.~\cite{DMCPV2022}, \(X=h_3\) and \(S=h_2\).

For the Heavy-resonant search, more generic benchmark models are considered, to represent a broader class of possible BSM models to which this analysis is sensitive.
The TRSM Benchmark 3 is used as a starting point, but only the cascade resonance decay (Figure~\ref{fig:intro:feynman:resonant_cascade}) is generated.
The widths of the \(X\) and \(S\) particles are set to \(1\%\) (\(20\%\)) of their masses, to define narrow (wide) width signal models.
These modifications lead to generic signal samples that no longer represent the TRSM.

The generic heavy-resonant signals are generated similarly to the other models, but using \MADSPIN~\cite{Artoisenet:2012st} for the \(S\rightarrow HH\) decay.
The Higgs bosons are decayed exclusively to \(b\bar{b}\).
At each width, 45 samples are produced, with \(550~\leq~\mX~\leq\SI{1500}{\GeV}\) and \(275~\leq~\mS~\leq~\SI{1000}{\GeV}\).

For all generated signal events, the detector response is simulated~\cite{SOFT-2010-01} using a fast parameterized simulation of the calorimeters~\cite{SIMU-2018-04}, and the full \GEANT~\cite{Agostinelli:2002hh} simulation for the other sub-detectors.
Events are processed using the same reconstruction software used for the data.
The effect of multiple interactions in the same and neighboring bunch crossings (\pileup) is modeled by overlaying the simulated hard-scattering event with inelastic \(pp\) events generated with \PYTHIA[8.186]~\cite{Sjostrand:2007gs} using the \NNPDF{2.3LO} PDF set~\cite{Ball:2012cx} and the A3 set of tuned parameters~\cite{ATL-PHYS-PUB-2016-017}.


%
%
%
\section{Event reconstruction}
\label{sec:objects}

%
For proton--proton collision events, primary vertices are reconstructed using at least two charged-particle tracks with \(\pt>\SI{500}{\MeV}\), measured in the ID~\cite{PERF-2015-01}.
The primary vertex with the largest sum of squared track momenta is assigned as the hard-scatter vertex.

Hadronic jets are reconstructed from particle flow objects~\cite{PERF-2015-09}, using the \antikt\ algorithm with a radius parameter of \(R=0.4\).
These are calibrated using a multi-step procedure as outlined in Ref.~\cite{JETM-2018-05}.
Jets are required to have \(\pt>\SI{20}{\GeV}\) and \(|\eta|<2.5\).
To suppress jets originating from \pileup, a neural-network based jet vertex tagger (\texttt{NNJVT}) is used, which is an updated version relative to the likelihood based algorithm described in Ref.~\cite{PERF-2014-03}.
Jets with \(20 < \pT < \SI{60}{\GeV}\) are required to satisfy a selection on the \texttt{NNJVT} with a \(0.88-0.99\) probability of correctly identifying hard-scatter jets depending on the \pT, as evaluated in \(Z(\rightarrow\mu\mu)+\text{jets}\) events with a mean of 33.7 interactions per bunch crossing.
Outside of this \pT\ range, no \texttt{NNJVT} requirement is applied.

Jets originating from \(b\)-quarks are identified using a DNN-based algorithm (\texttt{DL1d})~\cite{ATL-PHYS-PUB-2022-047,ATL-PHYS-PUB-2020-014,FTAG-2019-07}.
The identification requirement used selects jets with \(\pt>\SI{20}{\GeV}\) containing \(b\)-hadrons, with an efficiency of \(77\%\), and has a probability of \(0.38\%\) (\(15.5\%\)) of misidentifying light-flavor (charm) jets as determined in a sample of simulated \ttbar\ events.
The \(b\)-tagging efficiencies in simulated events are corrected to match those measured in data~\cite{FTAG-2018-01,ATLAS-CONF-2018-001,ATLAS-CONF-2018-006}.
Uncertainties related to the flavor-tagging selections are described in \Sect{\ref{sec:systs}}.

To account for energy lost to muons in semileptonic \(b\)-hadron decays, a momentum correction~\cite{HIGG-2016-29} is applied.
For this purpose, muons are reconstructed.
Muon candidates are found by matching ID tracks with either MS tracks or aligned individual hits in the MS and performing a combined track fit~\cite{MUON-2018-03}.
They are required to have \(\pt>\SI{4}{\GeV}\), \(|\eta|<2.5\), and to satisfy the \texttt{Medium} identification criteria defined in Ref.~\cite{MUON-2018-03}.
If any muons are within a cone of \(\Delta R = \min(0.4, 0.04 + 10/\pt^{\mu}~[\GeV])\) around the jet axis, their four-momentum is added to that of the jet.
Additionally, the energy in the calorimeter deposited by the muons (computed according to Ref.~\cite{PERF-2015-10}) is subtracted from the jet to avoid double-counting.

Corrections are applied to all signal MC simulated events to account for differences observed between the jet trigger efficiencies in data and simulation.
These corrections (scale factors, SFs) are calculated from the ratios of event-level jet trigger efficiencies in data and MC, which are in turn constructed from per-jet trigger efficiencies.
The efficiencies are estimated in \ttbar\ MC simulated events and data satifying a \ttbar-like event selection with at least one top quark decaying leptonically and collected using triggers requiring one muon and one jet to obtain an unbiased sample enriched in \(b\)-jets.
The \ttbar events were simulated using \POWHEGBOX[v2]~\cite{Frixione:2007nw,Nason:2004rx,Frixione:2007vw,Alioli:2010xd} at NLO with the \NNPDF[3.0nlo]~\cite{Ball:2014uwa} PDF set and A14 tune~\cite{ATL-PHYS-PUB-2014-021}.
\PYTHIA[8.230]~\cite{Sjostrand:2014zea} was used to model the parton shower, hadronization, and underlying event.
Separate SFs are estimated for the L1 and HLT, and for each jet as a function of \pt.

For each of the jets in the event selection, and for each data-taking year, the L1 trigger per-jet efficiencies are calculated.
The efficiency is defined as the fraction of the \(n^{\text{th}}\) \pT-ordered jets geometrically matched\footnote{The geometric matching requires \(\Delta R<0.4\) between the reconstructed and trigger-level jets.} to an L1 jet and passing the L1 trigger threshold in that year.
Similarly, the HLT per-jet efficiencies are calculated as the fraction of these L1-trigger-passing \(n^{\text{th}}\) jets that are geometrically matched to an HLT jet and pass the HLT trigger threshold in that year.
The L1 and HLT per-jet efficiencies are both calculated in bins of jet \pt.
These are combined to give overall event-level L1 and HLT jet trigger efficiencies, accounting for the combinatorial possibilities of selecting four jets that fired the trigger in events with more than four jets.
The L1 and HLT SFs are then calculated and multiplied to provide an overall SF for each event.


%
\section{Analysis strategy}
\label{sec:strategy}

\subsection{Event selection and strategy overview}
\label{sec:strategy:overview}

To be considered in the analysis, all data and MC simulated events must satisfy the trigger requirements defined in Section~\ref{sec:dataMC:data} and a set of preselection requirements, now described.
At least six jets are required, with at least four having \(\pT>\SI{40}{\GeV}\) in line with the trigger requirements.
At least four jets must be \btagged.

The analysis further categorizes events into three orthogonal regions, based on the multiplicity of \(b\)-tagged jets.
Denoted \fourb, \fiveb\ and \sixb\ respectively, these regions are defined to satisfy the preselection and required to have exactly four, exactly five, or at least six \btagged\ jets.
The \sixb\ region is enriched in the signal processes and is also referred to as the signal region (SR).
The \fiveb\ and \fourb\ regions are control regions (CRs) enriched in background processes and are used to derive the \sixb\ background estimate.
The overall acceptance times efficiency for the SM signal (for which only the \(H\rightarrow bb\) decay is considered) to fall into the \sixb\ region is 5.5\%.
For the \fourb\ and \fiveb\ regions the acceptance times efficiency is 27\% and 17\% respectively.

The general analysis strategy is the same for each of the three searches: Non-resonant, Resonant and Heavy-resonant.
In each case a DNN classifier is trained to discriminate between the signal models and the SM background, which is dominated by multijet production.
A profile likelihood fit is performed over the binned DNN score distribution to obtain the final results.

The kinematic properties of the reconstructed Higgs boson candidates provide important input features for the DNNs.
A mass-based pairing algorithm is used to assign jets to each of three Higgs boson candidates, as described in Section~\ref{sec:strategy:pairing}.

A fully data-driven background estimate is used for the analysis.
The approach assumes that the kinematic properties of the background processes are similar between events with four, five, or six identified \(b\)-jets in the final state.
Systematic uncertainties arising from this approach and its core assumption are discussed in \Sect{\ref{sec:systs}}.
The background estimate in the \sixb\ search region is obtained by extrapolating each DNN score distribution in data from the \fourb\ and \fiveb\ regions.
Both the \fiveb\ and \sixb\ regions are included in the fit.

Figure~\ref{fig:strategy:schematic} shows a diagram of the different regions in the analysis and how they are used in the final fit.
Each DNN score distribution is divided into three regions.
The ``High-Score'' region contains 90\% of the combined yields of the relevant signal samples, and provides the bulk of the analysis sensitivity in the fit.
The ``Low-Score'' region is enriched in background, and is used to constrain the background estimate in the fit as well as to derive systematic uncertainties on it.
The Low-Score \fourb\ data are used in the systematic uncertainty estimate but are not included in the final fits, preventing any double counting of degrees of freedom.
The ``Excluded'' region lies below the Low-Score region and is defined as the region in which the DNN score distribution extrapolation fails to model the observed \sixb\ data within one standard deviation of the statistical uncertainty.
Discarding the Excluded region reduces the size of the systematic uncertainties arising from the small differences in kinematics between the remaining Low- and High-Score regions.

\begin{figure}
\centering
\includegraphics[width=0.8\textwidth]{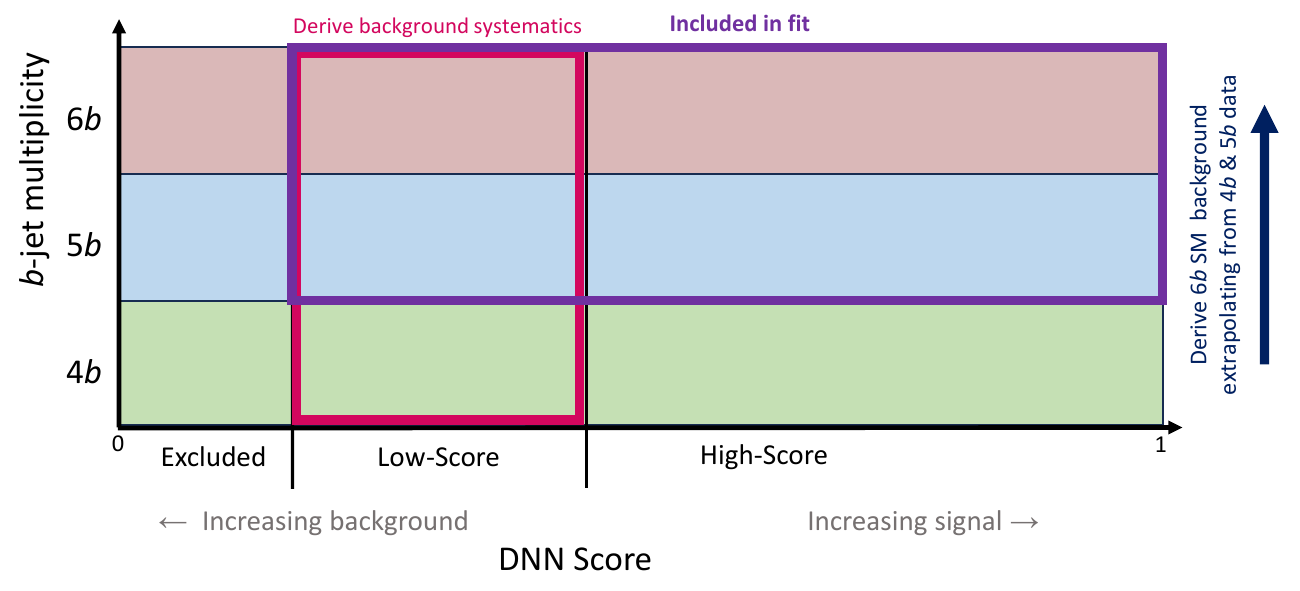}
\caption{A schematic diagram of the analysis strategy and regions. The strategy is the same for each DNN, although the precise selections defining the Excluded, Low-Score, and High-Score regions differ. The procedure to form the \sixb\ background estimate by extrapolating from data in the \fourb\ and \fiveb\ regions is described in Section~\ref{sec:backgrounds}.}
\label{fig:strategy:schematic}
\end{figure}

\subsection{Higgs boson reconstruction}
\label{sec:strategy:pairing}

To reconstruct the three Higgs boson candidates in signal events, which is useful to help discriminate between the signals and background, the six jets from the Higgs decays (labeled jets \(a-f\)) must be selected and paired.
Pairing is performed on all events satisfying the preselection requirements.
In the case of \sixb\ events, the six jets to pair are the six highest \pT\ \(b\)-tagged jets.
For \fourb\ (\fiveb) events, the six jets to pair are all of the \(b\)-tagged jets, plus the two (one) highest-\pT\ non-\(b\)-tagged jets.

The pairing algorithm used is defined by minimizing
\begin{equation}
|\mHiggs{1}-\SI{120}{\GeV}|+|\mHiggs{2}-\SI{115}{\GeV}|+|\mHiggs{3}-\SI{110}{\GeV}|,
\label{eq:pairing}
\end{equation}
over all possible permutations of jet pairs (for jets \(a-f\)) that satisfy these transverse-momentum constraints:
\begin{equation*}
\pT^{ab} > \pT^{cd} > \pT^{ef}, \textrm{where}~\pTX^{H1} = \pT^{ab},~\pT^{H2} = \pT^{cd}~\textrm{and}~\pT^{H3} = \pT^{ef}.
\end{equation*}
The numeric constants in Eq.~\ref{eq:pairing} are chosen based on the peaks of the \mHiggs{i} distributions in simulated signal events.
These differ from the observed Higgs boson mass of \SI{125}{\GeV} due to detector effects, energy lost to neutrinos from the \(b\)-hadron decays, and out-of-cone radiation.

The performance of the pairing algorithm is assessed using the pairing efficiency, which is defined as the fraction of signal events that were correctly paired out of all events where a correct pairing is possible.
Correct pairing may not be possible if the jets cannot all be geometrically matched to truth-level \(b\)-quarks from each Higgs boson decay. This can occur if any of the six jets falls out of the tracker acceptance (\(|\eta|>2.5\)), or if two \(b\)-quarks are close enough together that they are matched to the same jet.
The correct pairing is defined using a geometric matching between MC truth jets (reconstructed using the \antikt algorithm with $R = 0.4$ on stable, final-state particles from MC generators) and reconstructed jets.
Correct pairing is possible in 74--84\% of BSM signals, depending on \mX\ and \mS.
The pairing efficiencies for each signal model parameter plane are shown in Figure~\ref{fig:strategy:pairing}.
In scans over the \((\mX, \mS)\) space the pairing efficiency is calculated at each simulated mass point and a cubic interpolation using the nearest four points is used.
For the scan of the (\kthree, \kfour) space a parametric reweighting is used and therefore an arbitrarily smooth evaluation of pairing efficiency is achieved.
A correct pairing is possible in 81\% of SM signal events, and the pairing efficiency is 61\%.
The change in pairing efficiencies over the signal planes is a result of changing kinematics --- the higher the \(\pT\) of the Higgses, the easier the pairing.
For the SM-like and TRSM planes, the kinematic changes are due to the changes in interference between diagrams across the parameter space.

\begin{figure}
\centering
\subfloat[]{\label{fig:strategy:pairing:K3K4}\includegraphics[width=0.49\textwidth]{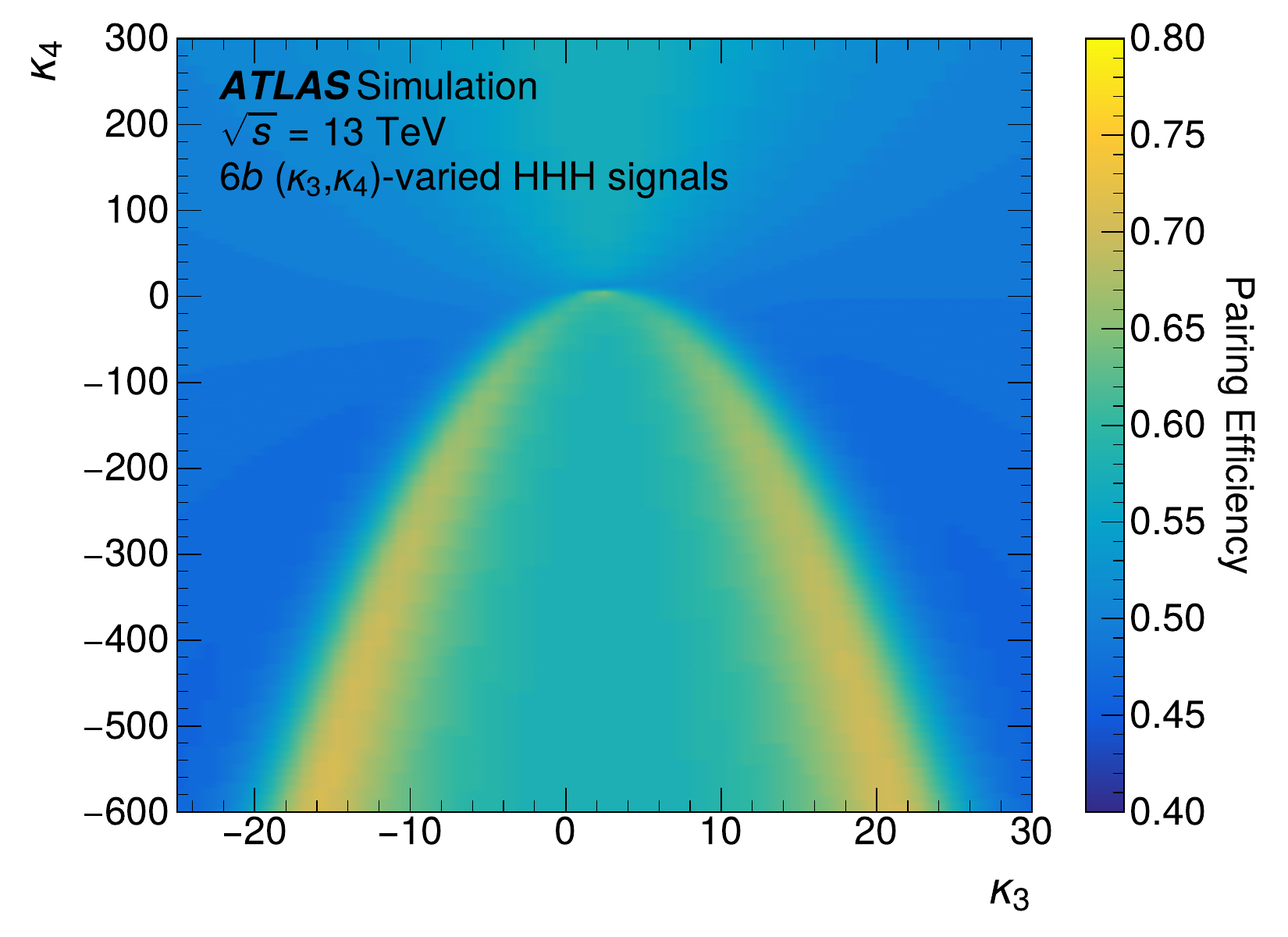}}
\subfloat[]{\label{fig:strategy:pairing:TRSM}\includegraphics[width=0.49\textwidth]{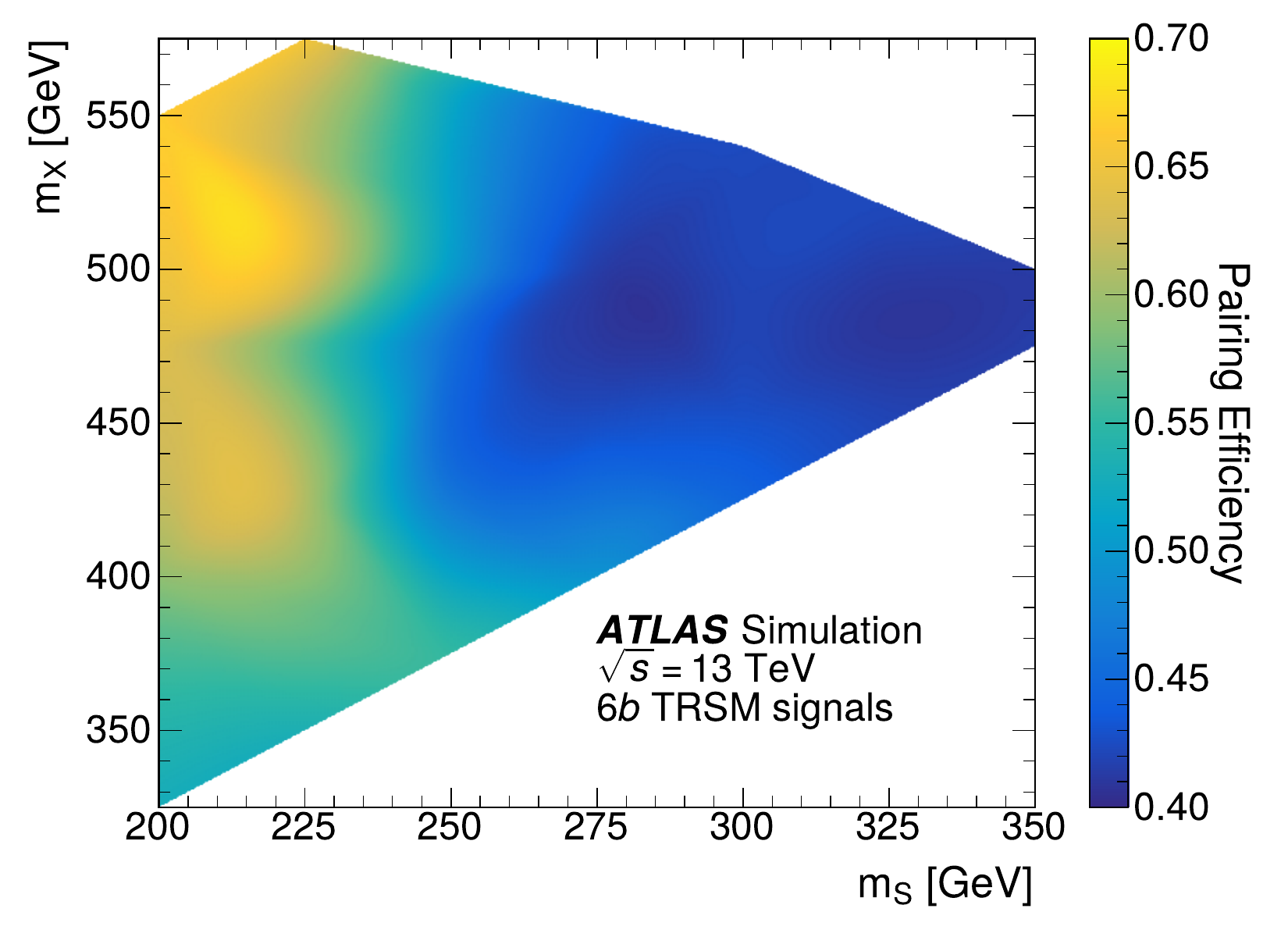}}\\
\subfloat[]{\label{fig:strategy:pairing:HeavyN}\includegraphics[width=0.49\textwidth]{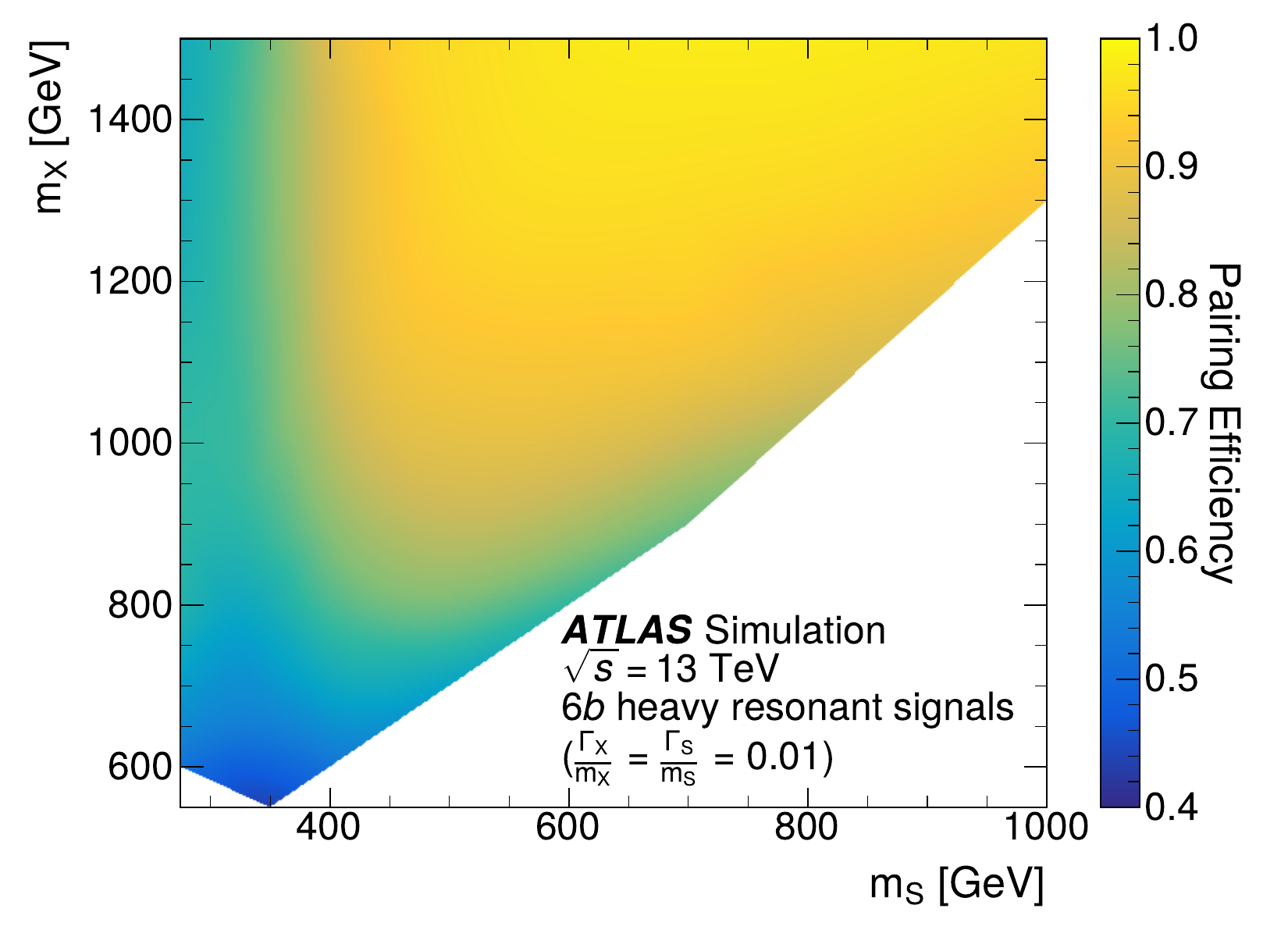}}
\subfloat[]{\label{fig:strategy:pairing:HeavyW}\includegraphics[width=0.49\textwidth]{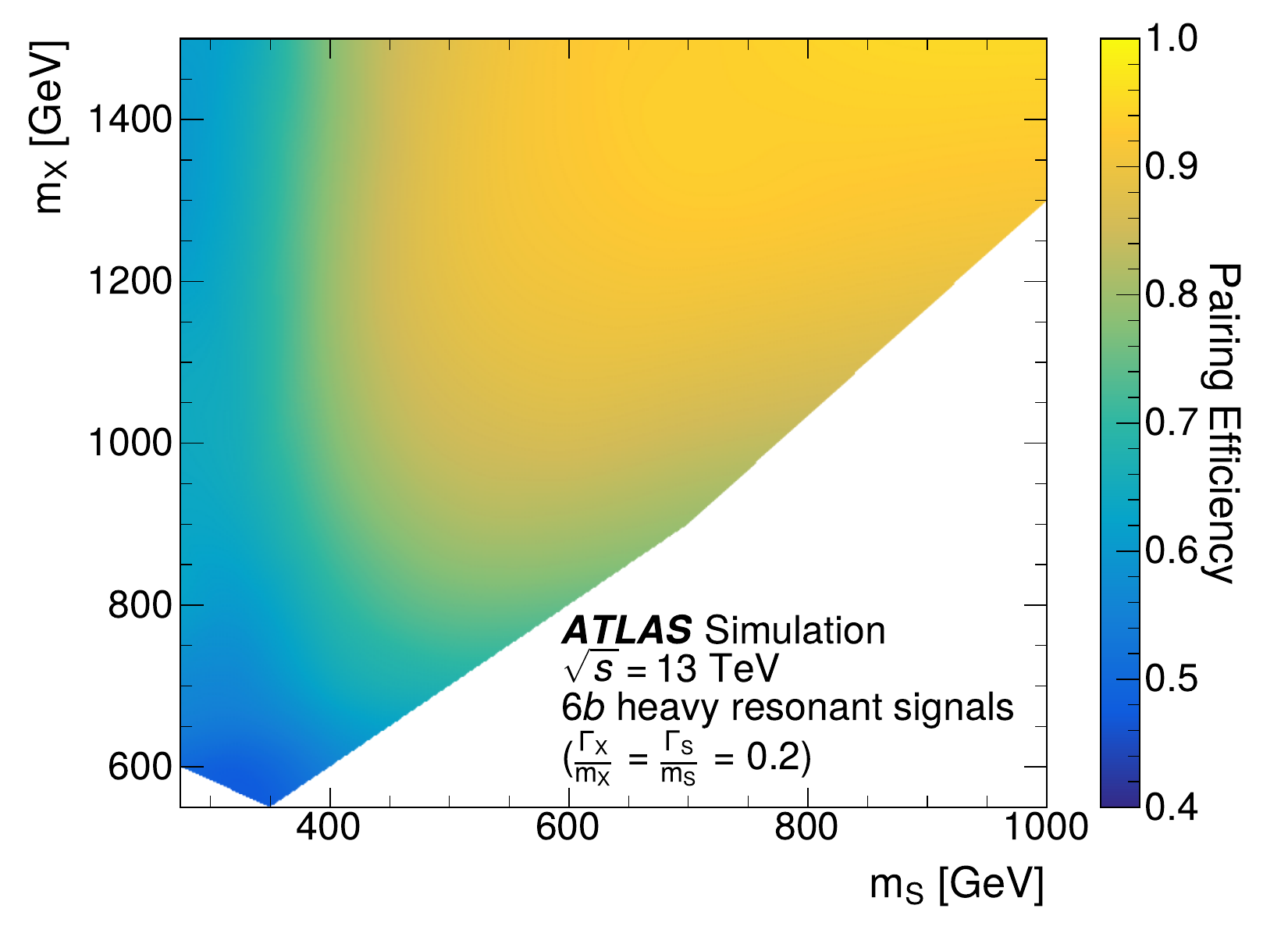}}
\caption{Jet pairing efficiencies over the parameter space for the (a) SM-like \((\kthree, \kfour)\) scan, (b) TRSM, and (c) narrow-width or (d) wide-width heavy resonance signals. The pairing efficiency is evaluated in the \sixb\ region when a correct pairing is possible --- that is, the six leading jets are geometrically matched to truth-level \(b\)-quarks. A cubic B\'ezier polynomial is used to interpolate across the plane. Areas of the \((\mX, \mS)\) phase space not covered by this search are shown in white.}
\label{fig:strategy:pairing}
\end{figure}

\subsection{Machine learning classifiers}
\label{sec:strategy:DNN}

A DNN is trained to classify events as signal- or background-like for each search: Non-resonant (nonresDNN), Resonant (resDNN) and Heavy-resonant (heavyresDNN).
The resulting classifier score distribution for each search is used as the final analysis discriminant.

For all DNNs, data events in the \fiveb\ region (which form the basis of the background estimation, as is discussed in Section~\ref{sec:backgrounds}) are used as the background training sample.
Potential signal contamination in \fiveb\ data is considered to have a negligible impact for this purpose, as it has yields of \(O(1\%)\) of the \fiveb\ data yield, depending on the signal model.
All resonant TRSM signals are combined to form the resDNN signal training sample, and all heavy resonance signals (both narrow and wide width) are combined to form the heavyresDNN signal training sample.
Each simulated (\mX, \mS) point is equally weighted in the training samples.
For nonresDNN, the SM \HHH\ signal, which has similar kinematics to the non-resonant TRSM signals, is used as the signal training sample.

These signals were chosen to maximize training sample size and to group samples with similar enough kinematics for the DNN to perform well.
The performance is enhanced by assigning ``class weights'' to the background such that the effective yield matches the signal yields during training.
To avoid overtraining and benefit from the full set of available events, orthogonal training and testing/validation datasets are used, through a \(k\)-fold splitting procedure with \(k=2\)~\cite{kfold}, split by odd vs even event numbers.
Two DNNs are trained such that the network trained on one half of the events is applied to the other half of the events in the analysis.

For each DNN ten discriminating kinematic variables are chosen as input features.
The ten variables were chosen from a larger set of kinematic variables that have good shape separation between signals and backgrounds.
The ten chosen features have the highest Shapley values~\cite{Shapley} (a common measure to quantify how impactful an input feature is on the classifier output), and minimize absolute correlations between variables to reduce redundancy.
Variables that are correlated differently in signal and background samples are \textit{not} excluded as they provide discriminating power.
In addition variables are chosen to have minimal correlation with the \(b\)-jet multiplicity to reduce the size of the systematic uncertainties in the \(b\)-jet multiplicity extrapolation.

The input features for each DNN are listed and defined in Table~\ref{tab:strategy:vars}.
Aplanarity and sphericity variables capture differences in the global event shape and are described further in Ref. ~\cite{STDM-2011-33}.
Non-resonant and low-mass resonant \HHH~tend to produce more isotropically distributed jets, while higher mass \(X\) resonances tend to produce more planar events.
The distributions of the most powerful kinematic variable for each DNN are shown in Figure~\ref{fig:strategy:kinematics}.
Distributions for \fourb, \fiveb\ and \sixb\ data and example signal samples are shown.
Ratios of data with different \bjet\ multiplicities, including the double ratio (ratio \sixb/\fiveb\ over ratio \fiveb/\fourb) are additionally shown.
The double ratio is used to estimate the background modeling systematic uncertainties as described in \Sect{\ref{sec:systs}}.

\begin{table}
\centering
\caption{Summary of the input variables used in each DNN. Check marks denote which input is used for each DNN.}
\begin{tabular}{p{1.9cm}p{8.7cm}p{1.1cm}p{1.1cm}p{1.1cm}}
\toprule
Variable & Definition & nonres & res & heavyres \\
\midrule
\mH-radius & Distance between the event (\(\mHiggs{1}, \mHiggs{2}, \mHiggs{3}\)) vector and expected vector for signal events, \((120, 115, 110)~\SI{}{\GeV}\). & \checkmark & & \checkmark \\
\midrule
\mHiggs{1}\ & Reconstructed mass of the highest \pT\ Higgs boson candidate. & \checkmark & & \checkmark \\
\midrule
\RMSmjj\ & Root-mean-squared (RMS) of the invariant mass of all 15 possible jet pairs. & \checkmark & & \checkmark \\
\midrule
\RMSdRjj\ & RMS of the angular separation between all 15 possible jet pairs. & \checkmark & \checkmark & \checkmark \\
\midrule
\RMSeta\ & RMS of the pseudo-rapidity of the Higgs boson candidates. & \checkmark & & \checkmark \\
\midrule
Skewness \dAjj\ & Skewness of \(\cosh(\Delta \eta_{ik}) - \cos(\Delta \phi_{ik})\), where \(i, k\) are all 15 possible jet pairs. & & \checkmark & \\
\midrule
\HTsixj\ & Scalar sum of the \pT\ of the 6 jets selected to reconstruct the 3 Higgs boson candidates. & & \checkmark & \\
\midrule
\(\cos\theta\) & In the \((\mHiggs{1}, \mHiggs{2}, \mHiggs{3})\) coordinate system, \(\theta\) is the angle between the vector from the origin to the event's reconstructed mass of the Higgs boson candidates, and the vector from the origin to \((120, 115, 110)~\SI{}{\GeV}\). & & \checkmark & \\
\midrule
Aplanarity\(_{6\text{j}}\) & The fraction of \pT\ from the 6 jets selected lying outside the plane formed by the 2 highest \pT\ jets \cite{STDM-2011-33}. & \checkmark & \checkmark & \checkmark \\
\midrule
Sphericity\(_{6\text{j}}\)& Isotropy of the momenta of the 6 jets selected to reconstruct the 3 Higgs boson candidates \cite{STDM-2011-33}. & & \checkmark & \\
\midrule
Transverse Sphericity\(_{6\text{j}}\)& Isotropy of the \pT\ of the 6 jets used for Higgs reconstruction, within the \(x-y\) plane \cite{STDM-2011-33}. & \checkmark & & \\
\midrule
Sphericity & Isotropy of the momenta of all jets in the event \cite{STDM-2011-33}. & & & \checkmark \\
\midrule
\(\eta-\mHHH\) fraction & \( \frac{\sum_{i,k} 2 \pT^{i}* \pT^{k} * (\cosh(\Delta \eta(ik))-1)}{\mHHH^2} \) where \(i, k\) are all 15 possible jet pairs, and \mHHH\ is the reconstructed tri-Higgs invariant mass. & & \checkmark & \\
\midrule
\dRH{1}\ & Angular separation between the jets paired to form the highest \pT\ Higgs boson candidate. & \checkmark & \checkmark & \checkmark \\
\midrule
\(\Delta R_{H2}\) & Angular separation between the jets paired to form the second-highest \pT\ Higgs boson candidate. & \checkmark & \checkmark & \checkmark \\
\midrule
\(\Delta R_{H3}\) & Angular separation between the jets paired to form the lowest \pT\ Higgs boson candidate. &\checkmark & \checkmark & \checkmark \\
\bottomrule
\end{tabular}
\label{tab:strategy:vars}
\end{table}

\begin{figure}
\centering
\subfloat[]{\label{fig:strategy:kinematics:mHradius}\includegraphics[width=0.49\textwidth]{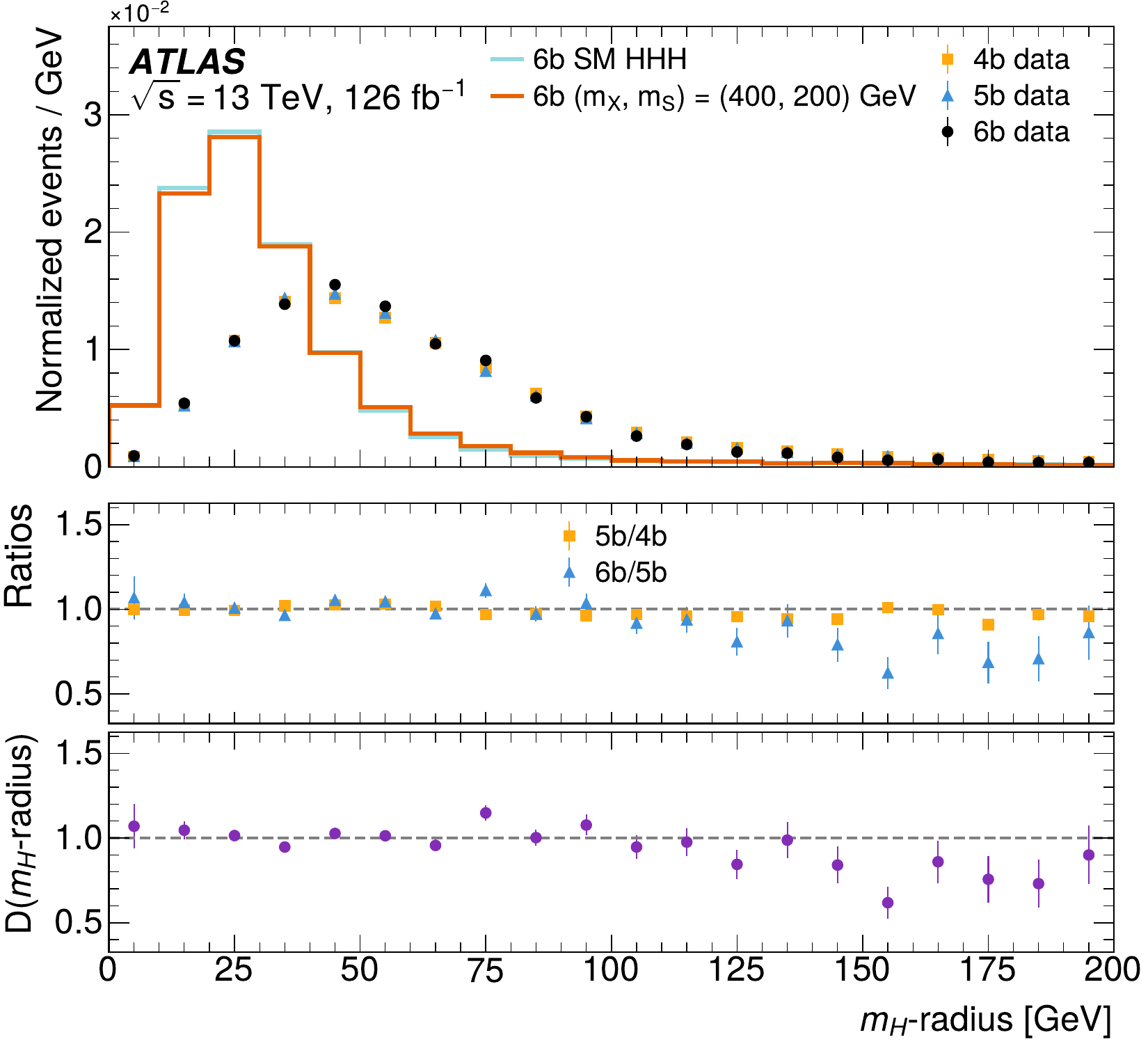}}
\subfloat[]{\label{fig:strategy:kinematics:massfraceta}\includegraphics[width=0.49\textwidth]{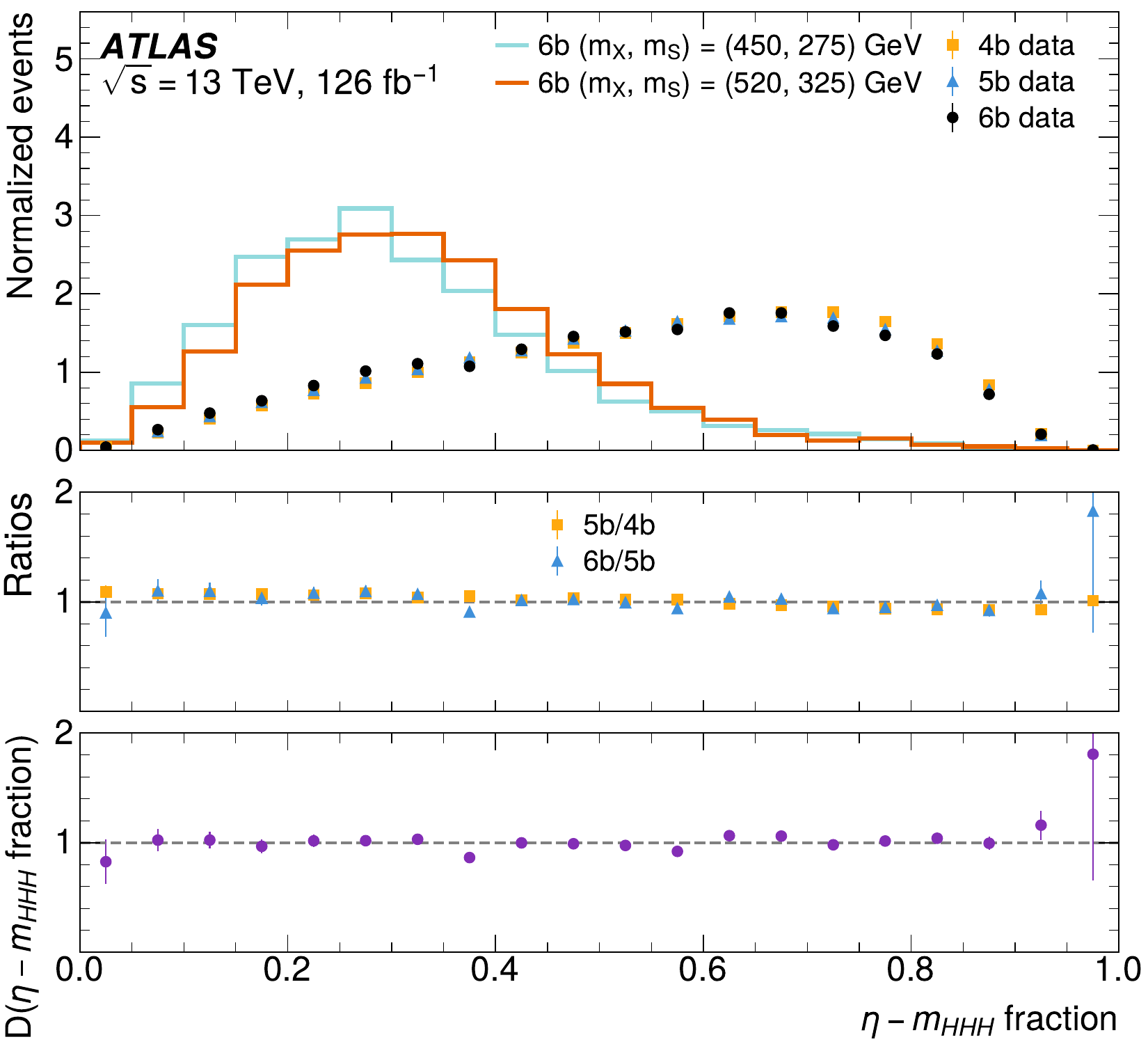}}\\
\subfloat[]{\label{fig:strategy:kinematics:dRH1}\includegraphics[width=0.49\textwidth]{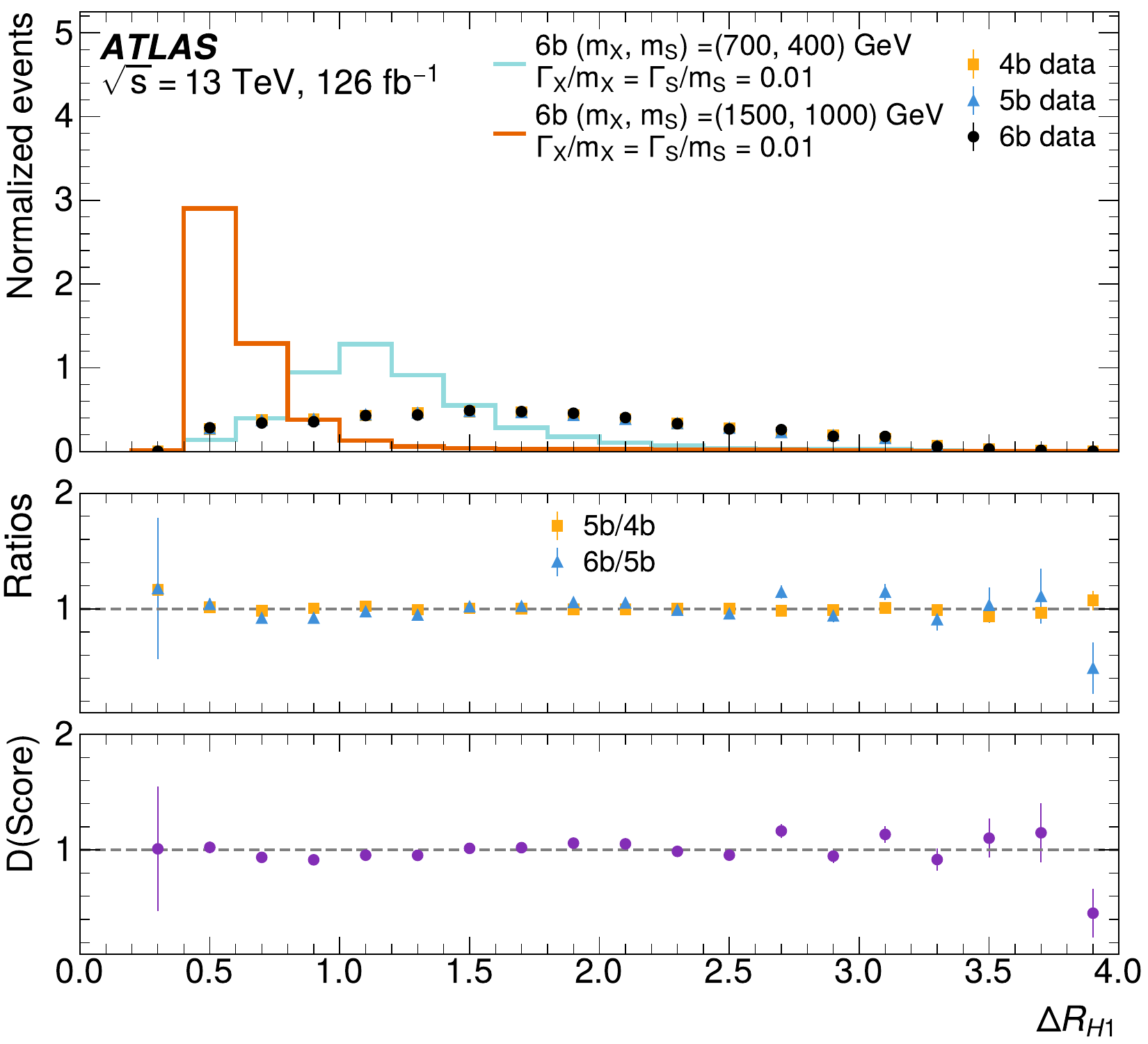}}
\caption{In the top panel of each plot, the distribution of the most discriminating variable used in each DNN training is shown: (a) \mH-radius for nonresDNN, (b) \(\eta-\mHHH\) fraction for resDNN, and (c) \dRH{1} for heavyresDNN. The distributions for the \fourb, \fiveb\ and \sixb\ data and example signal samples are shown, with all histograms normalized to unit area. In the middle panel of each plot, the ratios of data with different \bjet\ multiplicities are shown. Finally in the bottom panel, the ratio \sixb/\fiveb\ divided by the ratio \fiveb/\fourb\ --- \(D\) --- is shown for data; \(D\) is used to estimate the size of the systematic uncertainty in the \bjet\ extrapolation method as described in \Sect{\ref{sec:systs}}.}
\label{fig:strategy:kinematics}
\end{figure}

The DNNs are implemented using Keras~\cite{chollet2015keras} with a Tensorflow~\cite{tensorflow2015-whitepaper} backend.
A standard binary cross-entropy loss function is used, with rectified linear unit (ReLU) activation \cite{FukushimaRelu,HintonRelu}, an Adam~\cite{kingma2017adam} optimizer, and dropout regularization \cite{dropout} between layers.
The hyperparameters were optimized to maximize the sensitivity of the analysis.
These include three hidden layers with 24 hidden nodes per layer, a learning rate of 0.01 (0.005), a batch size of 128 (256), and a dropout rate of 0.1, for resDNN (nonresDNN).
For heavyresDNN, this includes four hidden layers with 24 hidden nodes per layer, a learning rate of 0.005, a batch size of 256, and a dropout rate of 0.1.

All DNN score distributions span from zero to one.
For nonresDNN, the Low-Score region starts above 0.100 (defined such that below this value the background DNN score distribution extrapolation fails to model the observed \sixb\ data within one standard deviation), and the High-Score region (containing 90\% of the combined signal yield) above 0.552.
For resDNN, the Low-Score region starts above 0.05, and the High-Score region above 0.367.
For heavyresDNN, the Low-Score region starts above 0.100, and the High-Score region above 0.790.


%
\section{Background estimation}
\label{sec:backgrounds}

As introduced in \Sect{\ref{sec:strategy}}, the dominant background process for a final state of \(\geq 6\) \(b\)-quarks is QCD multijet production.
The background is modeled via a data-driven approach using the data with four and five \bjets to predict the background yields with at least six \bjets.
The assumption of this method is that the kinematic properties of the background do not significantly change relative to \bjet multiplicity, as is observed in Figure~\ref{fig:strategy:kinematics} for the most important kinematic features used in the DNNs.
The assumption is further validated by comparing the ratio of \fourb\ and \fiveb\ data evaluated exclusively in the Low- and High-Score regions.
Therefore, background-enriched regions in DNN score and \bjet multiplicity are used to construct the \sixb\ background model and its associated systematic uncertainties in the search region.
The feature selection strategy (described in \Sect{\ref{sec:strategy:DNN}}) minimizes the difference between the shape of the DNN score distribution in \fourb\ and \fiveb\ data and therefore reduces the size of the associated uncertainties, which are discussed in \Sect{\ref{sec:systs}}.

Both the Low- and High-Score regions for \fiveb\ and \sixb\ data are included in the fit, as discussed in \Sect{\ref{sec:strategy}}.
The Low-Score region is used to estimate the background shape uncertainties, as discussed in \Sect{\ref{sec:systs}}.
The Excluded region tends to have large shape differences between \fourb, \fiveb\ and \sixb\ data and is not used for the remainder of the analysis.
The removal of the Excluded region is crucial to ensure that the analysis operates in a region where the kinematic properties are not highly dependent on \bjet\ multiplicity, reducing the size of the systematic uncertainties and ensuring that by construction the assumptions of the method remain valid in the Low- and High-Score regions.

Data in the \fourb\ and \fiveb\ Low- and High-Score regions are used to model the expected background contribution in the \sixb\ Low- and High-Score regions, respectively.
For each DNN an extrapolation across \bjet\ multiplicities is performed in the DNN score to model the expected background contribution in the \sixb\ region:
\begin{equation}\label{eq:bkg_model}
N_{i}^{6b\text{, predicted}} =\muNF \cdot B_{i} \cdot \left(\frac{(N^{5b}/N^{4b})_{i}}{N^{5b}/N^{4b}} \right),
\end{equation}
where \(N\) is the observed number of events (except when explicitly noted as in \(N_{i}^{6b\text{, predicted}}\)), \(B\) are a set of unconstrained bin-by-bin parameters fit from the \fiveb\ region, the superscript denotes the region (\fourb, \fiveb, or \sixb), and the subscript \(i\) denotes the \(i^\text{th}\) bin in the DNN score distribution.
The normalization on the background model is \muNF, which is fit in situ to the data as described in the next section.
In the case where there is no signal contribution, \(\muNF=N^{6b}/N^{5b}\) and \(B_i=N_i^{5b}\).
However, when the signal strength is non-zero \(\muNF<N^{6b}/N^{5b}\) and \(B_i<N_i^{5b}\).

The inclusion of the term \((N^{5b}/N^{4b})_{i}/(N^{5b}/N^{4b})\) in \Eqn{\ref{eq:bkg_model}} extrapolates the shape of the DNN score distribution as a function of \bjet\ multiplicity.
Such a shape extrapolation is observed to produce better modeling in the Low-Score region.


%
\section{Statistical interpretation}
\label{sec:stats}

Test statistics are built from the profile likelihood ratio (PLR) following the prescription for discovery of a positive signal and one-sided upper limits detailed in Ref.~\cite{Cowan:2010js}.
The likelihood model assumes independent Poisson-distributed event counts in each bin of the \fiveb\ and \sixb\ DNN score distributions in the Low-Score and High-Score regions, which are included in the fit.
Nuisance parameters (NPs) describing the effect of systematic uncertainties in the signal (\(\vec\phi\)) are fully correlated between the \fiveb\ and \sixb\ regions.
An orthogonal set of NPs \(\vec\theta\) describes the background extrapolation from the \fiveb\ to \sixb\ region.
The full expression of the parameterized likelihood model of the observed yields is:

\begin{equation}
\begin{split}
\mathcal{L}(\vec{N}^{\fiveb,\sixb},\vec{S}^{\fiveb,\sixb}|\vec{\theta},\vec{\phi},\vec{\gamma}^{\fiveb,\sixb},\musig,\muNF) = & \prod_{i \in \text{bins}} \left[ \text{Pois}\left(N_i^{\sixb} \middle\vert \muNF f_i(\vec{\theta})B_i + \musig w_i^{\sixb}(\vec{\phi})\gamma_i^{\sixb}S_i^{\sixb}\right) \right.\\
&\left. \times~ \text{Pois}\left(N_i^{\fiveb} \middle\vert B_i + \musig w_i^{\fiveb} (\vec{\phi})\gamma_i^{\fiveb}S_i^{\fiveb}\right) \right] \\
&\times \prod_{m \in \text{bkg. systs}} \frac{1}{\sqrt{2\pi}} e^{-\theta_m^2/2}\\
&\times \prod_{k \in \text{sig. systs}} \frac{1}{\sqrt{2\pi}} e^{-\phi_k^2/2}\\
&\times \prod_{i \in \text{bins}}\left[ P_\Gamma\left(\nu_i^{\fiveb} \middle\vert \gamma_i^{\fiveb}\nu_i^{\fiveb}\right)P_\Gamma\left(\nu_i^{\sixb} \middle\vert \gamma_i^{\sixb}\nu_i^{\sixb}\right) \right].\\ \label{eq:likelihood}
\end{split}
\end{equation}

The first two Poissonian terms describe the constraints from the observed data \(\vec{N}\).
Superscripts \fiveb\ and \sixb\ are used throughout to distinguish the \fiveb\ CR from the \sixb\ SR.
Signal leakage into the \fiveb\ region is considered via \(S_i^{\fiveb}\).
\(B_i\) are an unconstrained set of bin-by-bin yields fit from the CR and \(\musig\) is the signal strength and the parameter of interest (for the limits on the coupling modifiers \kthree\ and \kfour, the signal expression is altered such that \kthree\ and \kfour\ become two parameters of interest in the fit).
The precise manner in which \(\vec\phi\) and \(\vec\theta\) change each bin of the yields is encapsulated in \(w_i(\vec\phi)\) and \(f_i(\vec\theta)\), respectively.
The nominal value of \(f_i(\vec\theta)\) is the ratio of the DNN score shapes for the \fiveb\ and \fourb\ data, \(((N^{5b}/N^{4b})_{i})/(N^{5b}/N^{4b})\), as appears in \Eqn{\ref{eq:bkg_model}}.
Therefore, \(f_i(\vec\theta)\) performs the DNN score shape extrapolation to the \sixb\ region.
The uncertainties in the extrapolation are assessed using alternative shapes as discussed in \Sect{\ref{sec:systs}}.
The nominal value of \(w_i(\vec\phi)\) is unity, and it describes the way theoretical and experimental uncertainties both change the shape and normalization of the signal.

The third and fourth terms (products over \(m\) and \(k\)) describe the Gaussian constraints for the NPs.
The observed MC signal yields \(\vec{S}\) are multiplied by nuisance parameters \(\vec{\gamma}\) in the final term to account for the limited MC sample size.
The generalization of the Poisson distribution to continuous variables \(P_\Gamma\) is used to represent the MC statistical uncertainty by constraining \(\vec{\gamma}\).
The \(\vec{\nu}\) are constants constructed to provide the correct uncertainty sizes in the constraint term; they can be viewed as ``effective'' numbers of MC events, accounting for event weighting.

The choice of binning is different for each of the three DNNs.
The resDNN score distribution is fit from 0.05--1.0 in 19 bins of equal width 0.05.
The heavyresDNN score distribution is fit from 0.1--1.0 with equally spaced bins of 0.05.
Likewise, the nonresDNN score is fit in the region 0.1--1.0 with equal bins of width 0.05, except for the final bin which is subdivided into two equally sized bins of width 0.025.
Dividing the final bin in this search increases sensitivity.
The choice of binning is a trade-off between sensitivity and the ability to model systematic uncertainties in the data-driven background shape, as discussed in the next section.


\FloatBarrier

\section{Systematic uncertainties}
\label{sec:systs}

The systematic uncertainties considered include experimental uncertainties, uncertainties in the shape of the data-driven background modeling of the DNN score distribution, and theoretical modeling uncertainties in the simulated signal samples.

The dominant systematic uncertainty in this search is the uncertainty in the shape of the background DNN score distribution.
As described in \Sect{\ref{sec:backgrounds}}, the background model extrapolates the shape of the DNN score distribution as a function of the \bjet multiplicity, with the normalization of the background model \(\muNF\) determined in situ by the fit to the observed data.
Due to the data-driven nature of the background estimate, the shape extrapolation uncertainty is the only systematic uncertainty that affects the background prediction.
The shape extrapolation will agree with the \sixb\ data perfectly when the DNN score shape differences between the \sixb\ and \fiveb\ regions agree with the shape differences between \fiveb\ and \fourb.
However, this cannot be assumed, and the size of the non-closure in the extrapolation method in the signal-depleted Low-Score region is used to estimate the systematic uncertainty in the extrapolation method.
The double ratio \(D(v)\) is introduced, which compares the shape differences between the \sixb\ and \fiveb\ regions and the \fiveb\ and \fourb\ regions (each themselves ratios) as a function of the DNN input variable \(v\):

\begin{equation}
D(v) = \left(\frac{(N^{6b}/N^{5b})(v)}{N^{6b}/N^{5b}} \right)_{\text{Low-Score}} \div \left(\frac{(N^{5b}/N^{4b})(v)}{N^{5b}/N^{4b}} \right)_{\text{Low-Score}}
\end{equation}

where the ratios in the numerators are functions of the DNN input variable \(v\) in the Low-Score region.
The double ratio \(D(v)\) is thus the non-closure in the multiplicative extrapolation method in the Low-Score region as a function of \(v\).
The binned quantity \(D_i(v)\) is \(D(v)\) averaged over the events in bin \(i\) of the DNN score distribution.
In this context \(i\) runs over both Low- and High-Score regions, thereby applying the observed non-closure in the Low-Score region as a systematic uncertainty in the High-Score region.

Non-closure in the extrapolation method is quantified by deviations from \(D_i(v)=1\).
The nominal background model (\Eqn{\ref{eq:bkg_model}}) is re-weighted by the value of \(D_i(v)\) to produce a set of systematically varied models, with each of the 10 DNN inputs producing a different template.
The variations are considered fully correlated between each bin \(i\) of the DNN.
This method does not rely on any assumption that \(D_i(v)=1\).
Rather, it assumes that deviations from \(D_i(v)=1\) due to background processes are similar in the Low-Score and High-Score regions.
This assumption is validated in the background-enriched \fourb\ and \fiveb\ regions.
It is observed that the ratios \((N^{5b}/N^{4b})(v)\) evaluated in the Low- and High-Score regions are compatible within statistical fluctuations, an agreement which has been reinforced by removing the very low score events (Excluded region) from the analysis.
The multiplicative variations to the nominal background template from \(D_i(v)\) are symmetrized about \(D_i(v)=1\) for up and down variations.

Only a subset of 10 possible shape variations (one for each of the 10 DNN input variables) are applied as systematic uncertainties.
A pruning procedure is used that removes shape variations that are expressible as linear combinations of other shape variations to avoid double counting.
The pruning procedure is to iteratively add new systematic variations (ordered by largest single bin deviation) until the set of excluded variations are well modeled by the set of included variations.
Specifically, a profile likelihood fit to each of the excluded variations is performed, with the only degrees of freedom being the included shape variations.
The procedure terminates once the maximum \(\chi^2\) of all of the fits to the excluded shapes is less than 1.
It is found that for each DNN, a subset of three shape variations is enough to model all 10 shapes sufficiently well.

\Fig{\ref{fig:systs:kinematicVariations}} shows the pruned subset of shape variations for each of the DNNs.
The shape variations are always smaller than the statistical uncertainty in the expected \sixb\ data.
For reference the \fiveb\ data without extrapolation is shown.
The difference between the observed \fiveb\ data and the extrapolated nominal \sixb\ template is on the same order as the derived shape systematic variations.

\begin{figure}
\centering
\subfloat[]{\label{fig:systs:kinematicVariations:nonres}\includegraphics[width=0.49\textwidth]{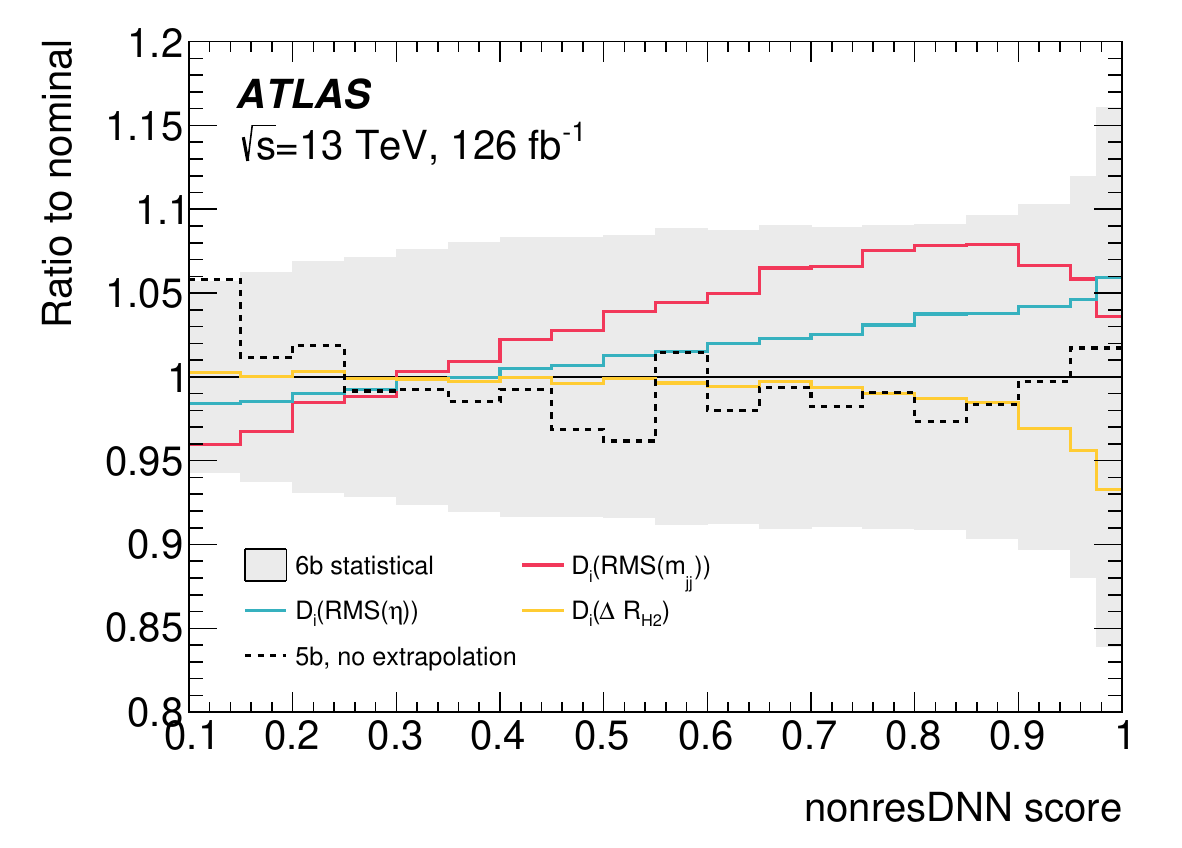}}
\subfloat[]{\label{fig:systs:kinematicVariations:res}\includegraphics[width=0.49\textwidth]{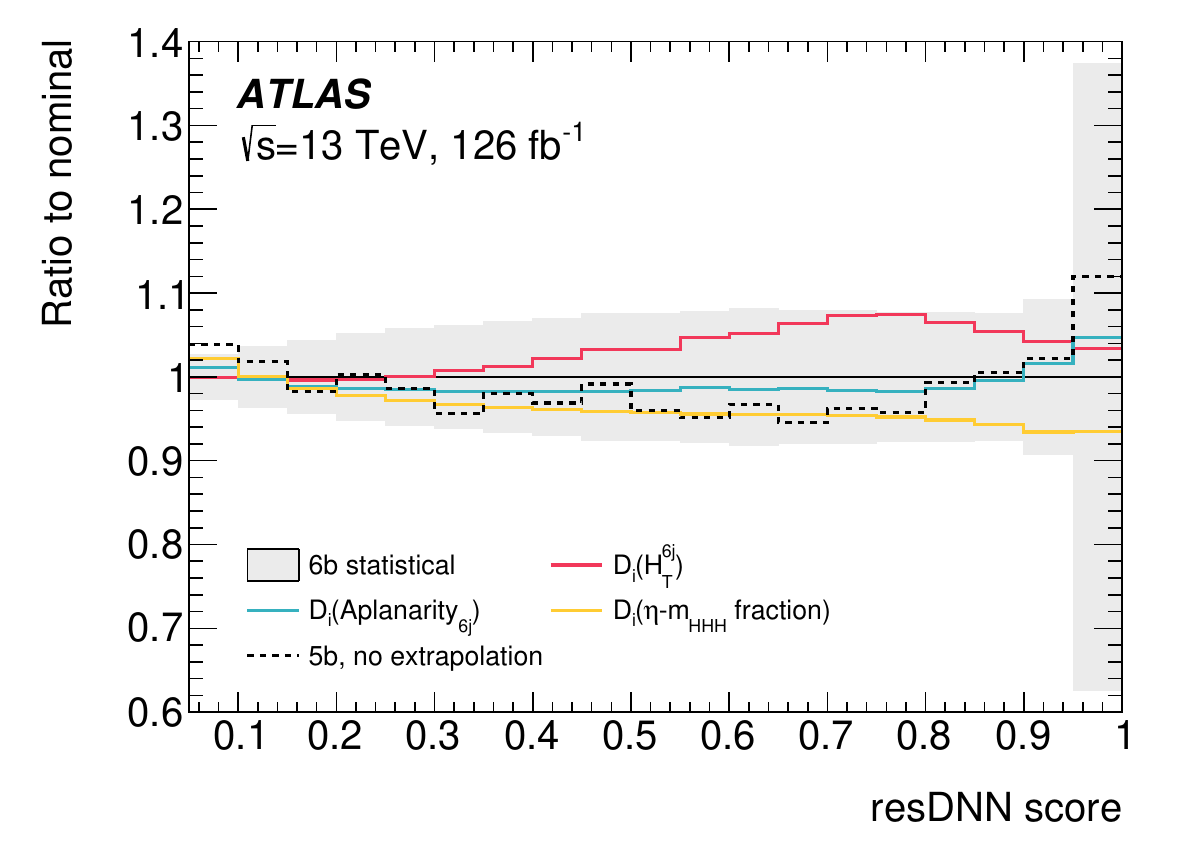}}\\
\subfloat[]{\label{fig:systs:kinematicVariations:hevres}\includegraphics[width=0.49\textwidth]{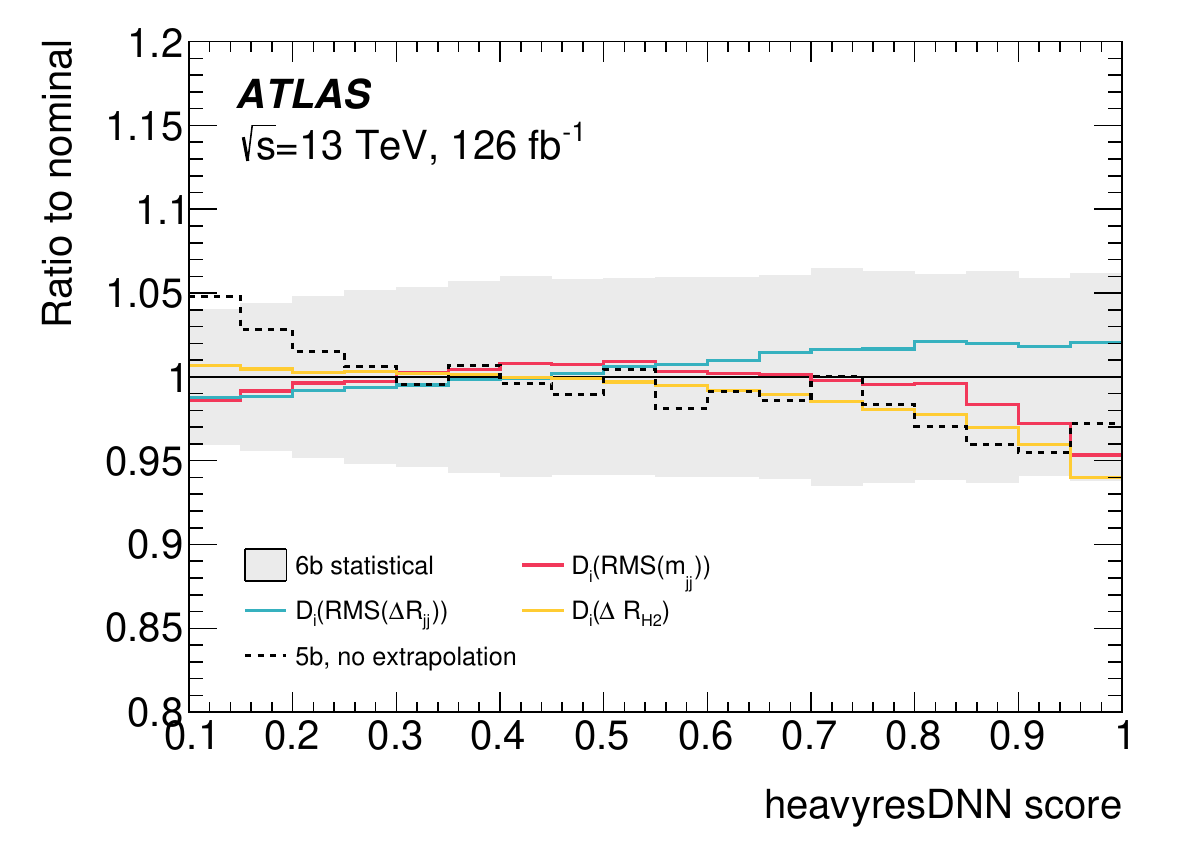}}
\caption{Ratios of the shape-varied models relative to the nominal background models (\(D_i(v)\)) for the selected input kinematics that survived pruning of the (a) nonresDNN (b) resDNN and (c) heavyresDNN. The gray band is the size of the statistical uncertainty in the expected \sixb\ background for reference. The dashed line is the shape of the \fiveb\ data with no extrapolation applied. The shape systematic uncertainties are all smaller than the expected \sixb\ background statistical uncertainty for each bin and are on the same order as the size of the extrapolation from \fiveb\ data. The binning matches exactly the scheme used for the PLR fit as described in \Sect{\ref{sec:stats}}.}
\label{fig:systs:kinematicVariations}
\end{figure}

Experimental uncertainties in luminosity, \pileup\ reweighting, jet energy scale and resolution, \texttt{NNJVT} efficiency, trigger efficiency, and \btagging efficiency are considered.
As the background model is entirely data-driven, these uncertainties only affect the simulated signal MC events.
The \pileup\ uncertainty is estimated according to the methods described in Ref. \cite{STDM-2015-05}.
Uncertainties in the jet energy scale and resolution, \texttt{NNJVT} efficiency, and \btagging efficiency are applied on a per-jet level and are estimated from calibrations and comparisons between different event simulation models \cite{PERF-2015-09,JETM-2018-05,PERF-2014-07}.
The uncertainty in the \texttt{NNJVT} efficiency is estimated following the technique in Ref. \cite{PERF-2014-03}.

The uncertainty in the trigger efficiency is estimated in data and MC using triggers requiring one muon and one jet so as not to bias the fully hadronic trigger efficiency measurement.
The uncertainty in the jet-level trigger SF is estimated as the difference between the nominal SF and a SF derived using \ttbar\ MC samples with different matrix elements and parton shower models.
One alternative sample uses an alternative matrix element next-to-leading-order (NLO) matching scale (pThard), and another uses \texttt{Herwig7}~\cite{Bellm:2015jjp} in place of \texttt{Pythia8} for the parton showering.
The event-level SF applied is built from the jet-level SFs considering all possible combinations of which jets may have fired the trigger.

A 5\% uncertainty is applied to the signal MC samples to cover the differences between the \(b\)-tagging applied at trigger level (\texttt{MV2c10} for 2016 and \texttt{MV2c20} for 2017--2018)~\cite{TRIG-2018-08} and at reconstruction level (\texttt{DL1d})~\cite{ATL-PHYS-PUB-2022-047,ATL-PHYS-PUB-2020-014,FTAG-2019-07}.
Given that an event has satisfied the offline selection of six \(b\)-jets, it has a larger than \(99\%\) chance of satisfying the trigger's online \(b\)-tagging.
Therefore the 5\% uncertainty is much larger than the trigger inefficiency in the region most sensitive to the presence of a signal.

Modeling uncertainties in the signal MC event simulations are also considered and are calculated using an identical procedure for all signal models considered.
To estimate the uncertainty due to higher-order terms in the perturbative expansion of the cross-section a 7-point variation in renormalization and factorization scales is used.
Both scales are either kept at their nominal values or varied independently up and down by a factor of 2 and 0.5, while the schemes in which one scale is varied up and the other is varied down are excluded for a total of seven possible choices of renormalization and factorization scales.
The uncertainty applied to the signal MC is an envelope around the seven variations.
An alternative value of the strong coupling constant \(\alpha_S\) is also used and the difference is taken as an uncertainty.
A systematic uncertainty in the choice of PDF is applied as the standard deviation of 100 variations in the \NNPDF{3.0NLO} PDF set.

\Tab{\ref{tab:systematic_impact}} summarizes the impact of each set of systematic uncertainties in the expected upper limit.
Each row shows the difference between the upper limit with all sources of uncertainty considered and the upper limit with a subset of the systematic uncertainties excluded.
Each cell shows the range of impacts for the various mass hypotheses \mX\ and \mS\ considered for each of the three DNNs.
The systematic uncertainties in the background shape are the dominant sources of systematic uncertainty, changing the expected limit by 14 to 45\% depending on the model and BSM masses \mX~and \mS.
Flavor tagging, jet reconstruction and trigger efficiency uncertainties are the sub-leading sources of systematic uncertainty, with an impact of up to 10\% on the expected upper limit.

\begin{table}
\centering
\caption{Summary of the impact on the analysis of different sources of systematic uncertainty. The impact is estimated as the absolute value of the difference between the expected upper limit with all sources of uncertainty included and the upper limit with a subset of systematic uncertainties excluded. In the left-hand column, more heavily indented rows are a sub-category of the less indented row above.}
\label{tab:systematic_impact}
\begin{tabular}{lcccc} \toprule
\multirow{2}{*}{Uncertainty source} 		& \multicolumn{4}{c}{Relative impact of systematic uncertainties [\%]} \\ \cline{2-5}
& SM-like	& TRSM non-resonant	& TRSM resonant		& Heavy resonance \\ \toprule
\textbf{All uncertainties}					& 24		& 20--46				& 33--42				& 24--53 \\ \midrule
\hspace{1em}\textbf{Experimental}			& 22		& 20--45				& 33--41				& 24--53 \\ \midrule
\hspace{2em}\textbf{Detector response}	& 7.4		& 6.6--14			& 16--24			& 4.1--15 \\
\hspace{3em}Luminosity and \pileup	& $<$1		& $<$1				& $<$1				& $<$1 \\
\hspace{3em}Flavor tagging			& 3.2		& 2.8--5			& 6.9--8.8			& 1.5--5.6 \\
\hspace{3em}Jet reconstruction		& 2.7		& 2.3--6.5			& 3.6--7.1			& 1.0--6.3 \\
\hspace{3em}Trigger efficiency		& 2.0		& 1.8--3.5				& 6--10				& 1.4--4.2 \\
\hspace{2em}\textbf{Background modeling}	& 16		& 14--36				& 18--30				& 20--45 \\ \midrule
\hspace{1em}\textbf{Theoretical}			& 1.5		& $<$1				& $<$1				& $<$1 \\ \midrule
\hspace{1em}\textbf{MC statistical}			& $<$1		& $<$1				& $<$1				& $<$1 \\ \midrule
\end{tabular}
\end{table}


\FloatBarrier

\section{Results}
\label{sec:results}
%
%

%
The distributions of the nonResDNN, resDNN, and heavyResDNN scores in the \(\sixb\) region after the profile likelihood fit described in Section~\ref{sec:stats} can be found in Figure~\ref{fig:results:postFitDNN}.
One representative signal point is overlaid on each score distribution, scaled to the same area as the background distribution.
For nonresDNN, the SM \(HHH\) signal and a non-resonant TRSM signal with \((\mX, \mS)=(400,200)~\SI{}{\GeV}\) are both shown in \Fig{\ref{fig:results:postFitDNN:nonResSM}} and \Fig{\ref{fig:results:postFitDNN:nonResTRSM}}, respectively.
In all cases, the observed data agree well with the background, and no significant excess is seen.
For the Resonant search no upward deviation from the SM prediction is observed.
A one-sided test statistic is used, and the best fit signal is precisely zero for all mass points considered.
For the Non-resonant search, the largest deviation is at \((\mX, \mS)=(550,200)~\SI{}{\GeV}\) with a local significance of \(0.19\sigma\).
For the Heavy-resonant search, the largest deviation is at \((\mX, \mS)=(1500,275)~\SI{}{\GeV}\) with a local significance of \(0.51\sigma\).

\begin{figure}[!h]
\centering
\subfloat[]{\label{fig:results:postFitDNN:nonResSM}\includegraphics[width=0.49\textwidth]{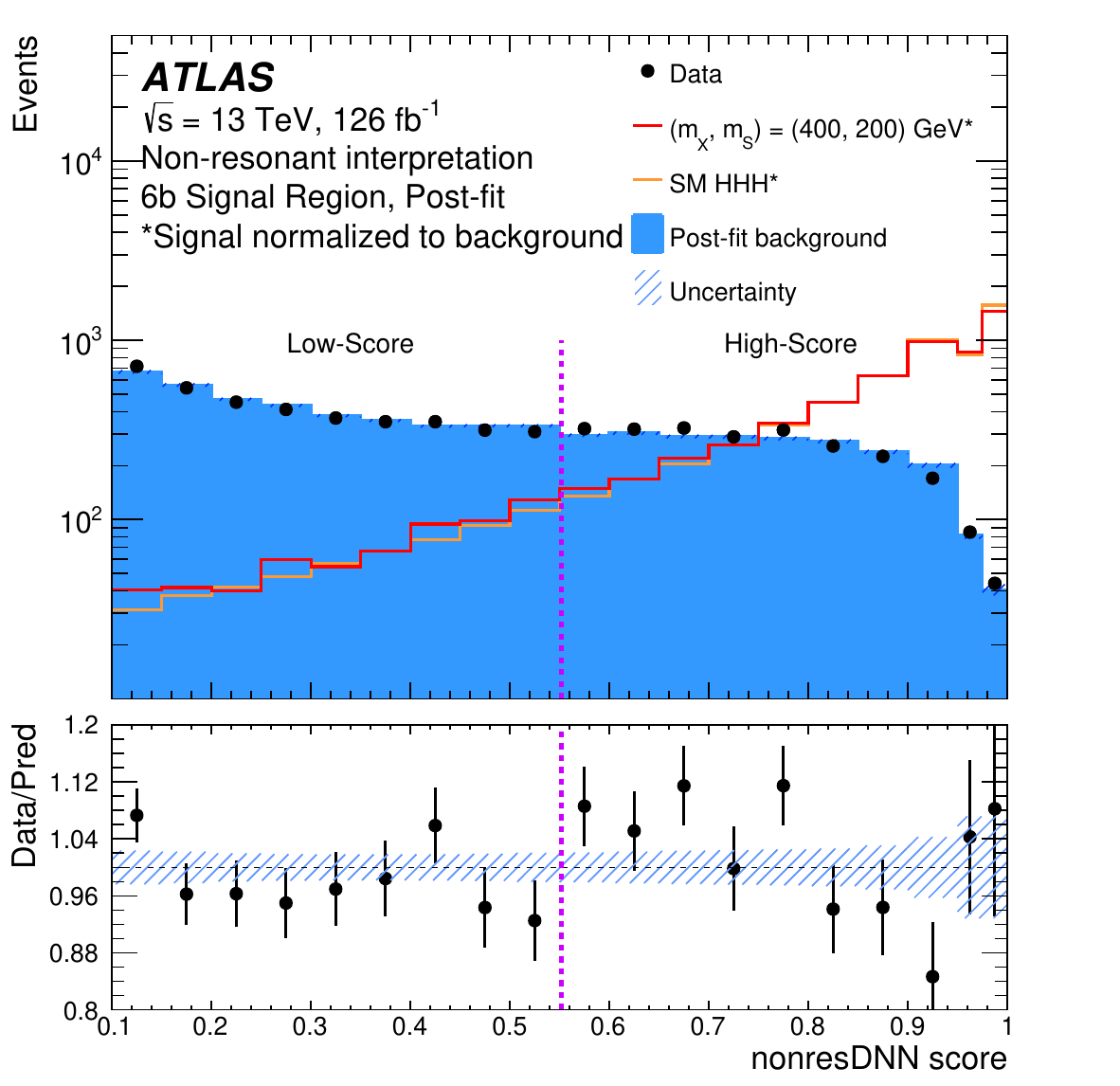}}
\subfloat[]{\label{fig:results:postFitDNN:nonResTRSM}\includegraphics[width=0.49\textwidth]{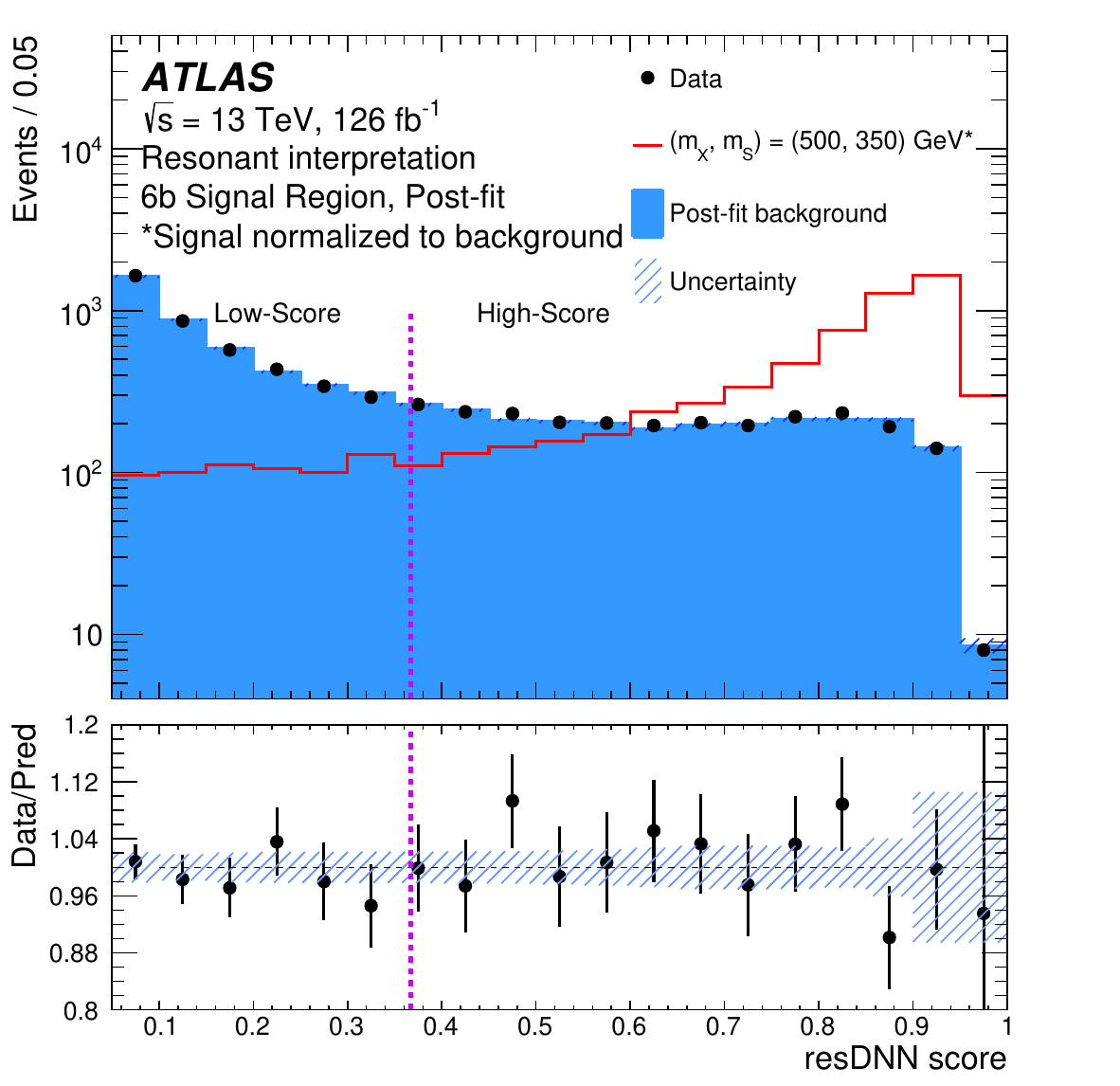}}\\
\subfloat[]{\label{fig:results:postFitDNN:Res}\includegraphics[width=0.49\textwidth]{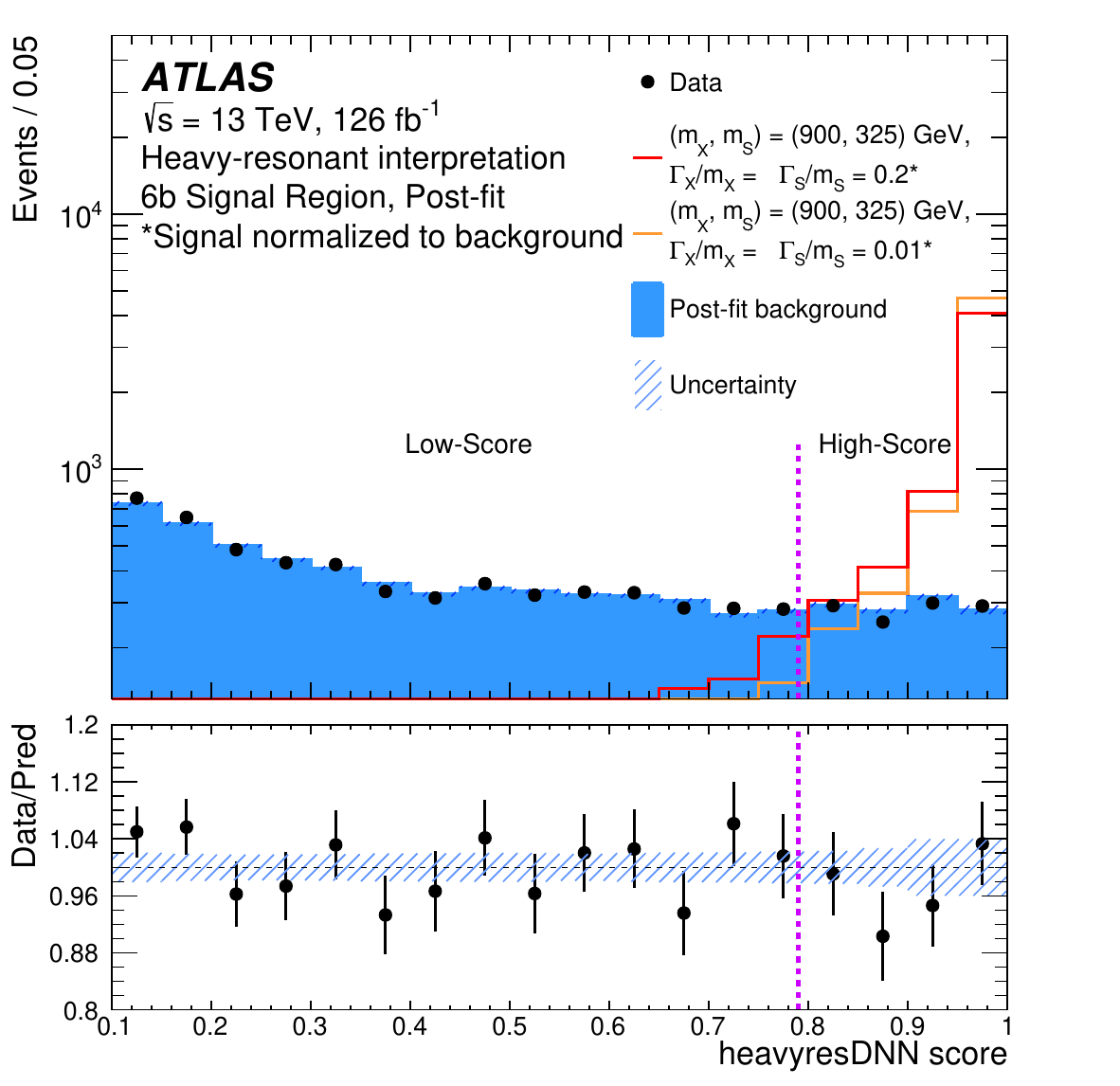}}
\caption{
Distributions of scores for (a) nonresDNN, (b) resDNN and (c) heavyresDNN in \sixb\ data and the background prediction after background-only fits to the observed data.
The boundary between the Low- and High-Score regions is shown by a dashed vertical line.
Benchmark signal models (normalized to the background) are overlaid in each case: (a) SM \(HHH\) signal and non-resonant TRSM signal with \((\mX, \mS)=(400,200)~\SI{}{\GeV}\), (b) resonant TRSM signal with \((\mX, \mS)=(500,350)~\SI{}{\GeV}\) overlaid, and (c) narrow- and large-width heavy resonances with \((\mX, \mS)=(900,325)~\SI{}{\GeV}\) overlaid.
The lower panel in each plot shows the ratio of the data and the post-fit background (Pred) in markers, with the uncertainty on the prediction shown as a hatched band centered on unity.}
\label{fig:results:postFitDNN}
\end{figure}

Upper limits for each signal model considered are placed on the allowed \(HHH\) production cross-section, at 95\% CL.
They are calculated using the \(CL_s\) method~\cite{Read:2002hq} with the asymptotic approximation~\cite{Cowan:2010js}.

For the TRSM and DM-CPV models, expected and observed cross-section upper limits are shown over the \((\mX,\mS)\) plane in Figures~\ref{fig:results:limitsTRSM:exp} and \ref{fig:results:limitsTRSM:obs}, respectively.
The \(\mS>\SI{250}{\GeV}\) region forms the Resonant search, where resDNN is used in the fit, whereas the \(\mS<\SI{250}{\GeV}\) region is part of the Non-resonant search, where nonresDNN is used in the fit.
The expected (observed) cross-section upper limit varies with \(\mX\) and \(\mS\), within the range of \(46-\SI{350}{fb}\) (\(48-\SI{310}{fb}\)).
The limits are evaluated at the simulated \((\mX, \mS)\) points as described in \Sect{\ref{sec:dataMC}}.
For all results in the analysis, a cubic B\'ezier polynomial is used to interpolate across the plane.

\begin{figure}[!h]
\centering
\subfloat[]{\label{fig:results:limitsTRSM:exp}\includegraphics[width=0.49\textwidth]{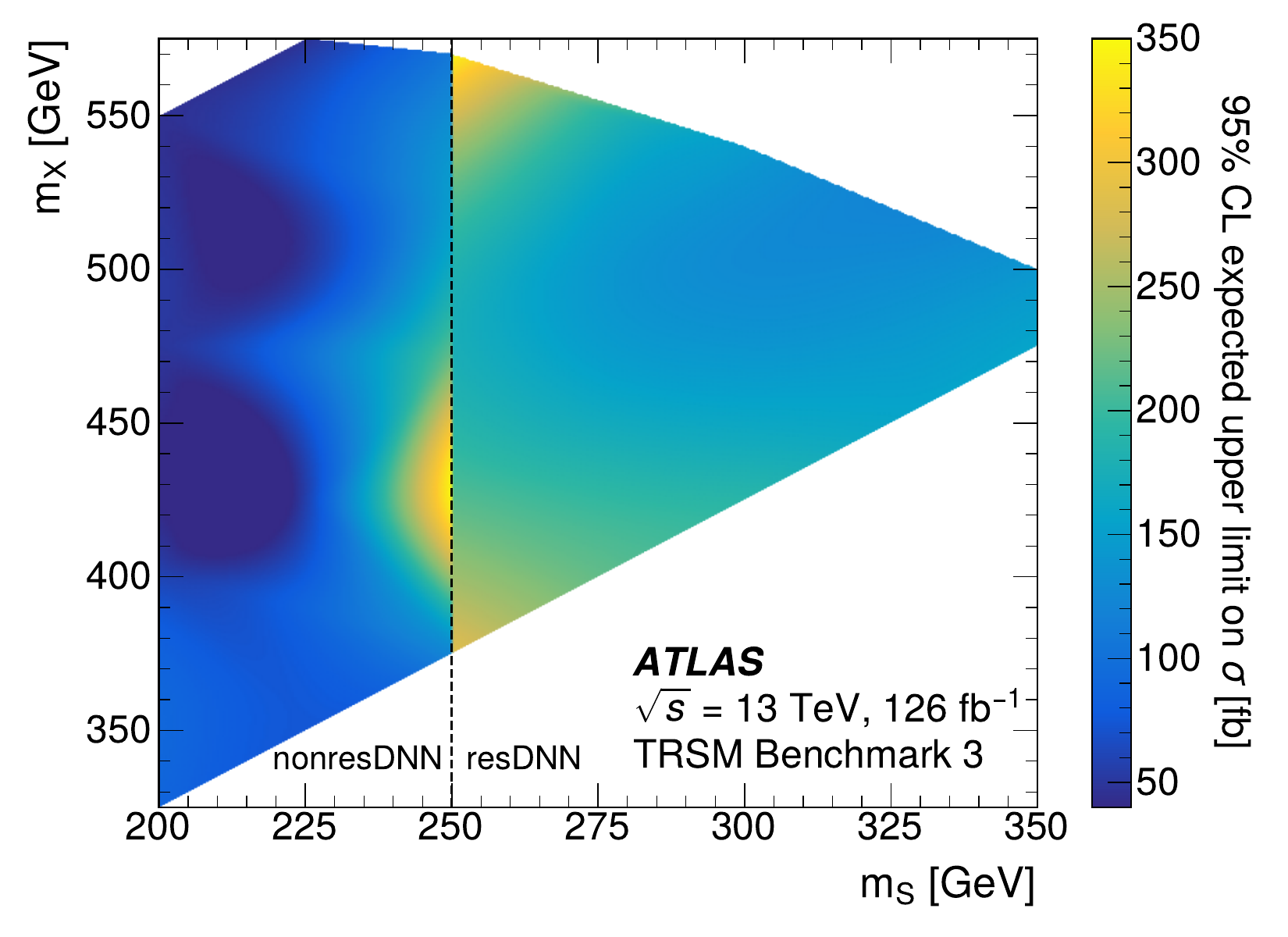}}
\subfloat[]{\label{fig:results:limitsTRSM:obs}\includegraphics[width=0.49\textwidth]{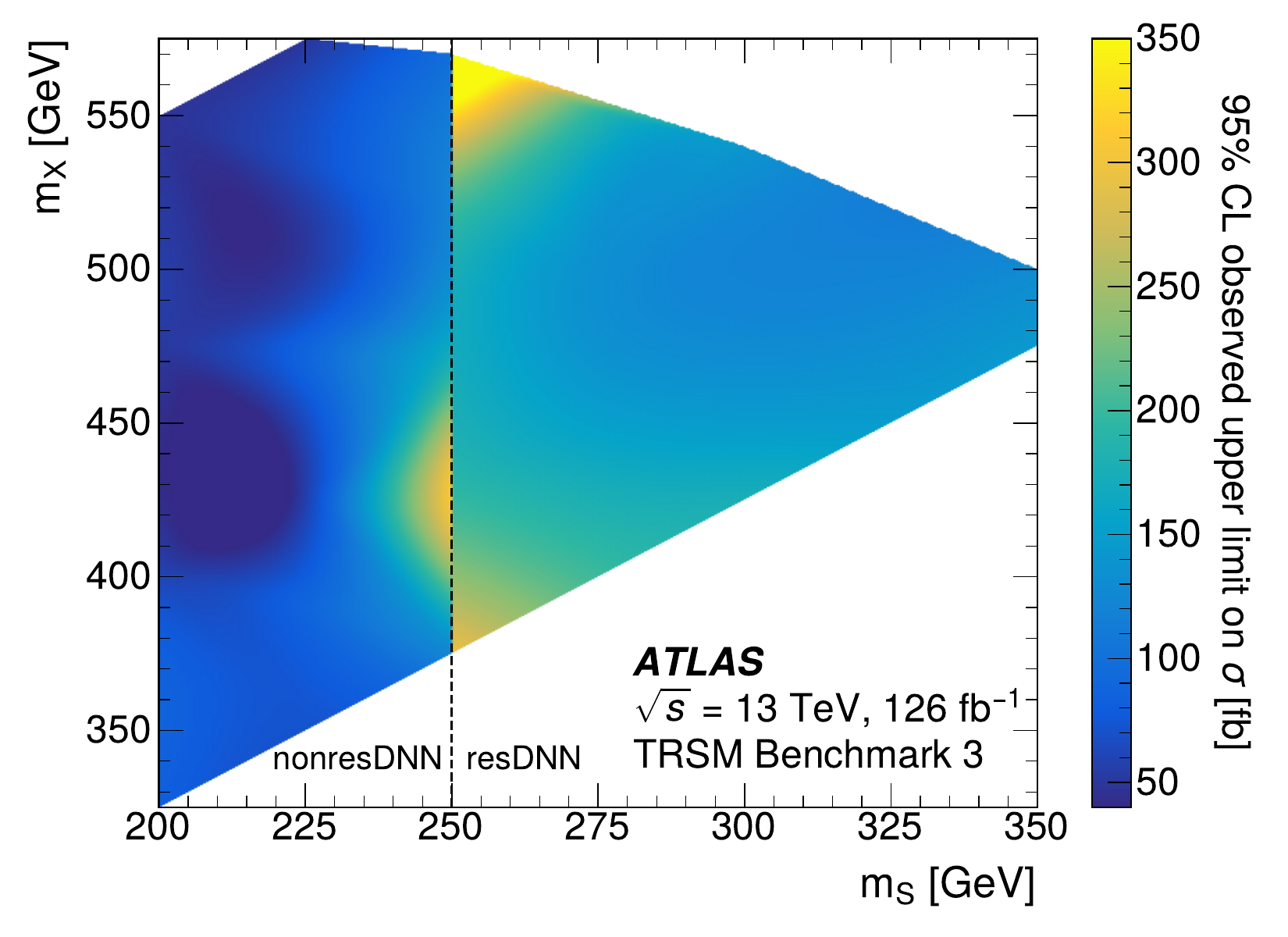}}
\caption{(a) Expected and (b) observed 95\% CL \(HHH\) cross section upper limits for the phase space within the perturbative unitarity bounds of the TRSM. While the points were generated under Benchmark point 3 of the TRSM, they can also be interpreted in the DM-CPV model. The white areas correspond to regions of the \((\mX, \mS)\) phase space not considered here. A cubic B\'ezier polynomial is used to interpolate across the plane.}
\label{fig:results:limitsTRSM}
\end{figure}

The cross-section upper limits for the Heavy Resonance search are shown in Figure~\ref{fig:results:limitsHeavy} in the \((\mX,\mS)\) plane.
Expected (observed) limits for the narrow heavy resonance signals are presented in \Fig{\ref{fig:results:limits:HeavyN:exp}} (\Fig{\ref{fig:results:limits:HeavyN:obs}}), and range between \(4.7-\SI{69}{fb}\) (\(5.7-\SI{38}{fb}\)).
Expected (observed) limits for the wide heavy resonance signals are presented in \Fig{\ref{fig:results:limits:HeavyW:exp}} (\Fig{\ref{fig:results:limits:HeavyW:obs}}), and range between \(5.2-\SI{53}{fb}\) (\(6.3-\SI{39}{fb}\)).

\begin{figure}[!h]
\centering
\subfloat[]{\label{fig:results:limits:HeavyN:exp}\includegraphics[width=0.49\textwidth]{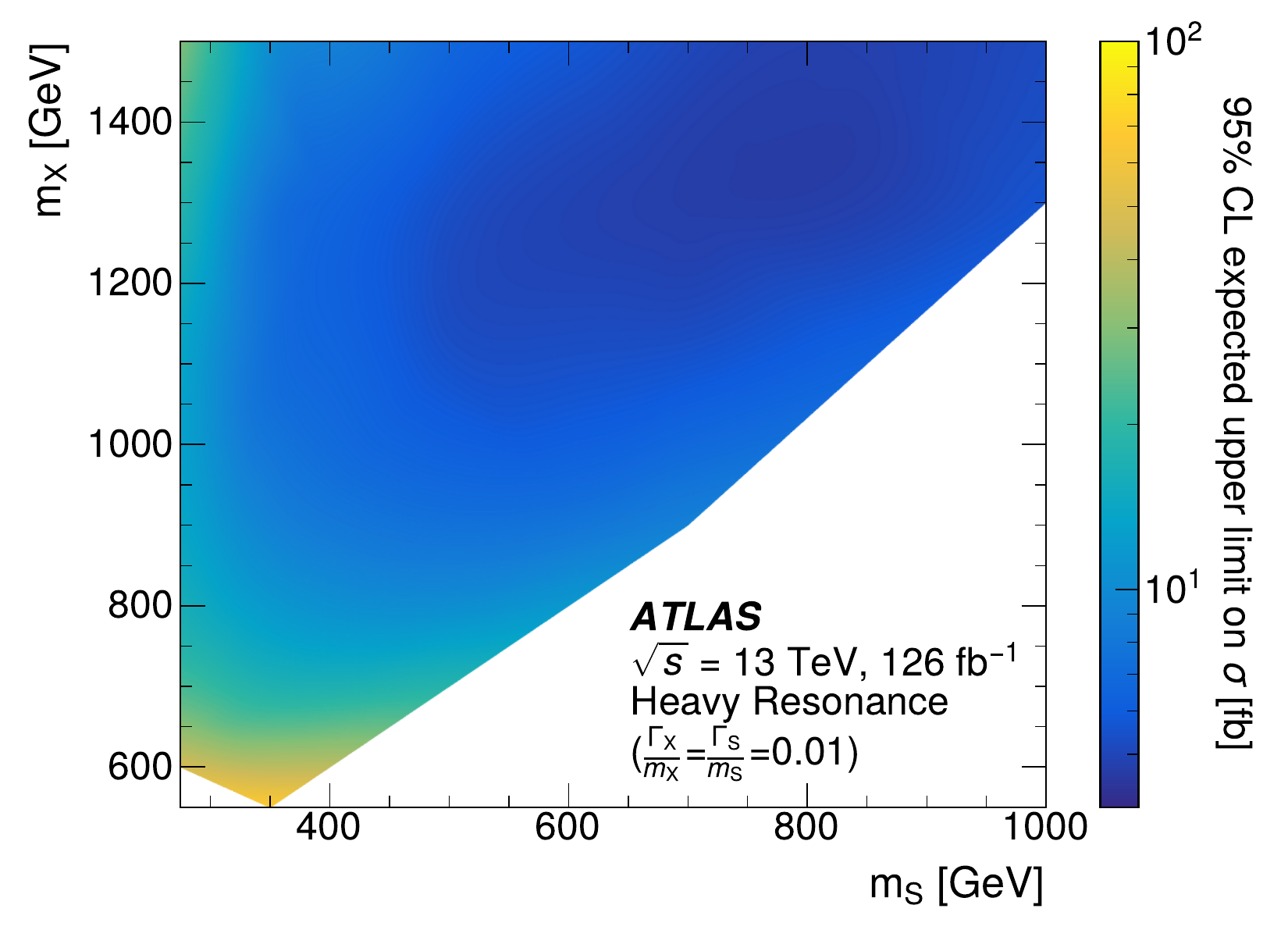}}
\subfloat[]{\label{fig:results:limits:HeavyW:exp}\includegraphics[width=0.49\textwidth]{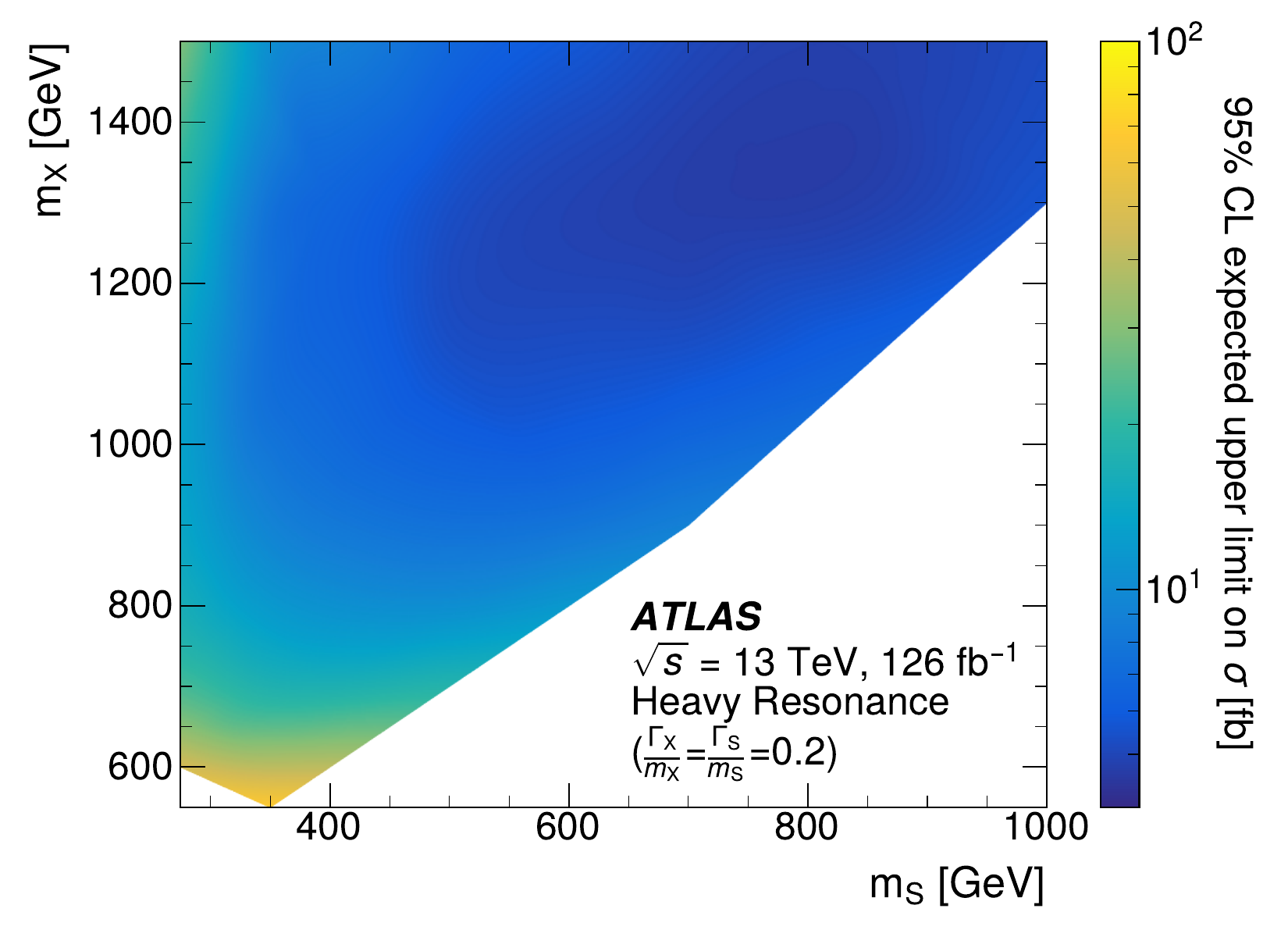}}\\
\subfloat[]{\label{fig:results:limits:HeavyN:obs}\includegraphics[width=0.49\textwidth]{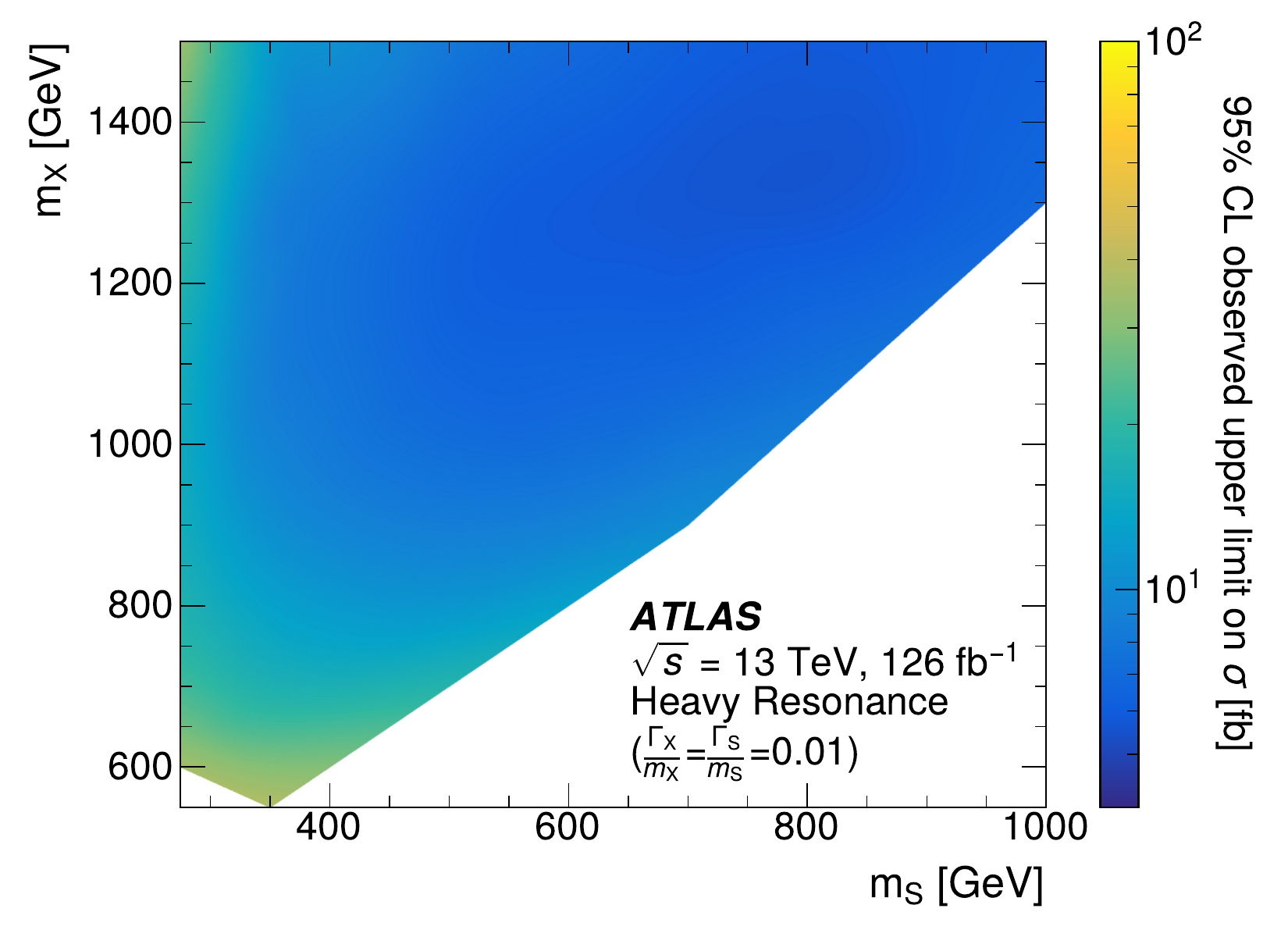}}
\subfloat[]{\label{fig:results:limits:HeavyW:obs}\includegraphics[width=0.49\textwidth]{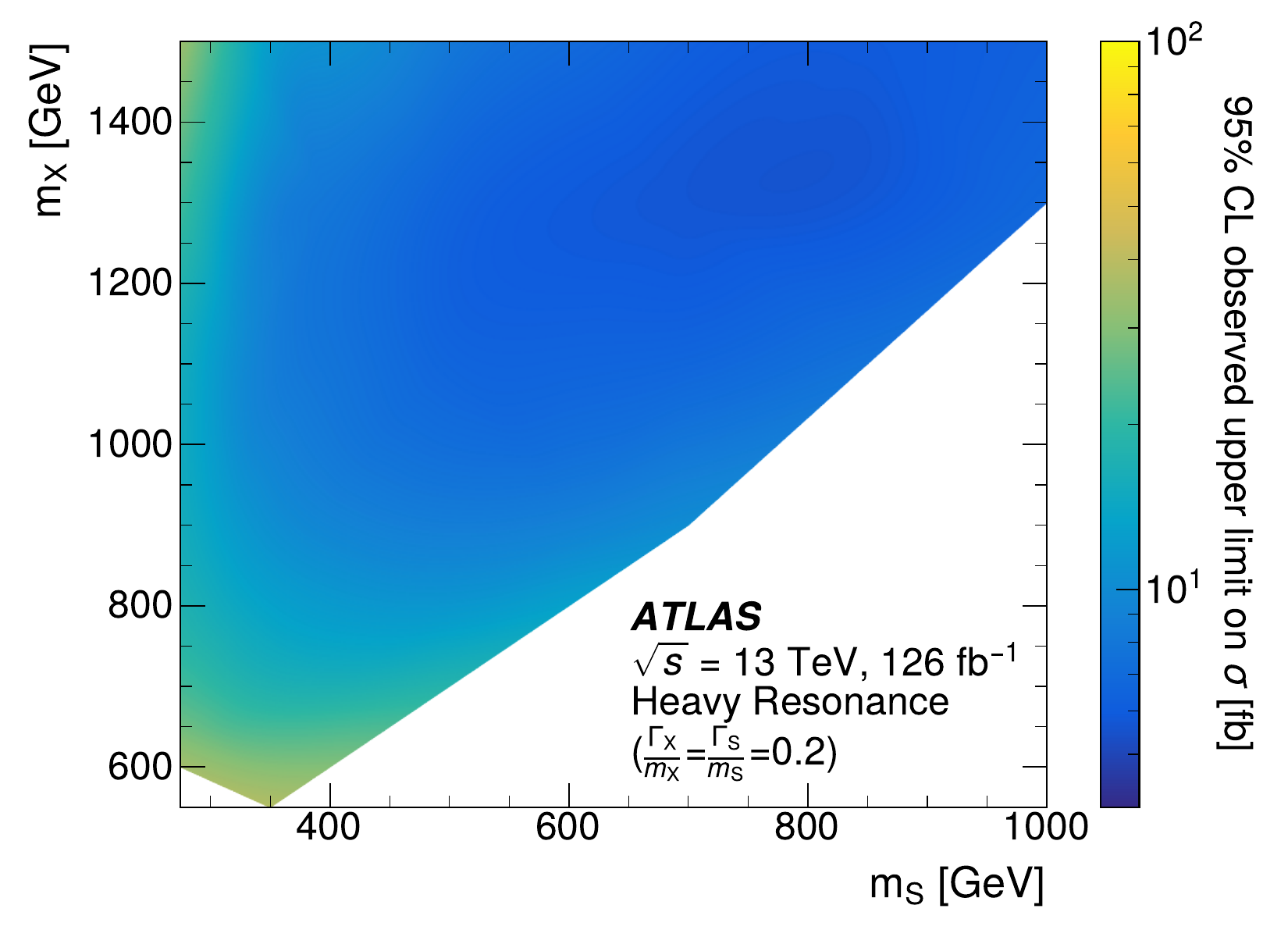}}
\caption{(a,b) Expected and (c,d) observed 95\% CL \(HHH\) cross section upper limits for the (a,c) narrow-width or (b,d) wide-width heavy resonance signals. The white areas correspond to regions of the \((\mX, \mS)\) phase space not considered here. A cubic B\'ezier polynomial is used to interpolate across the plane.}
\label{fig:results:limitsHeavy}
\end{figure}

\begin{figure}[!h]
\centering
\includegraphics[width=0.7\textwidth]{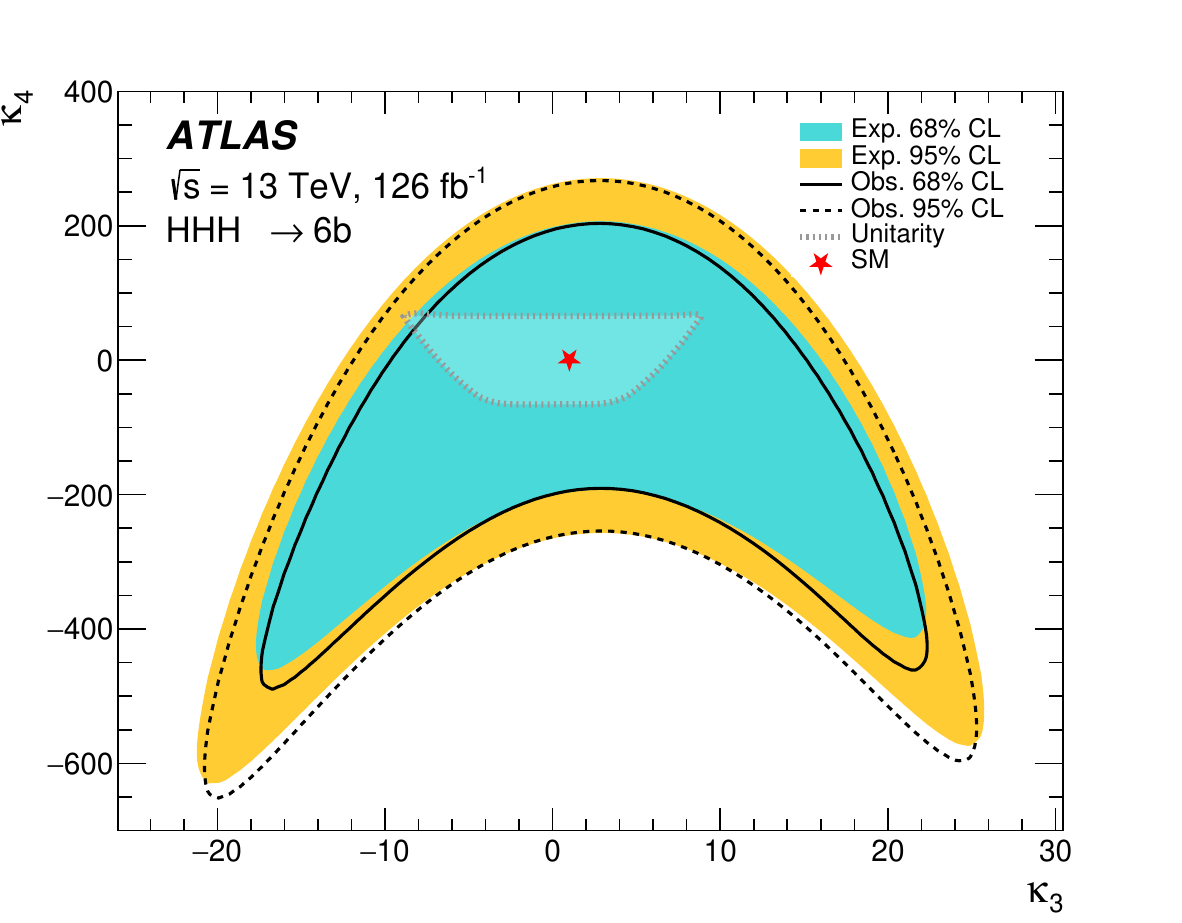}
\caption{Expected (filled regions) and observed (black solid and dashed lines) 95\% and 68\% CL constraints on \kthree\ and \kfour, the ratios of the Higgs tri-linear and quartic self-couplings to their predicted SM values. Unitarity limits, calculated in Ref.~\cite{Stylianou:2023xit}, are overlaid in the region bounded by the gray dashed line. The red star is the SM: \kthree=\kfour=1.}
\label{fig:results:limitskappa}
\end{figure}

Simultaneous limits on the coupling modifiers \kthree\ and \kfour\ are shown in \Fig{\ref{fig:results:limitskappa}}.
The gray dashed line shows the region where perturbative unitarity holds (provided that \kthree\ and \kfour\ are the only modifications to the SM), as calculated in Ref.~\cite{Stylianou:2023xit}.
At the 95\% CL none of the phase space inside the unitarity bounds is excluded.
Outside the unitarity bounds the kappa framework requires additional modification to preserve unitarity, and as such it is not recommended to interpret this result as excluding any relevant phase space in the kappa framework.
Rather, the scan of \kthree\ and \kfour\ serves as a benchmark of performance for future searches and projections, as well as other new-physics models which may produce similar phenomenology.

The observed (expected) 95\% CL upper limit on the signal strength for SM \(HHH\) production \(\mu=\sigma_{HHH}/\sigma_{HHH}^{SM}\) is 760 (750), corresponding to an observed cross-section upper limit of \(\SI{59}{fb}\) (\(\SI{59}{fb}\)).
Assuming \(\kfour=1\) then \(\kthree\) is restricted to be between \(-11\) and 17 at 95\% CL (both for expected and observed limits).
For comparison, a combination of previous di-Higgs searches and single Higgs production constraints limited \(\kthree\) to be between \(-0.4\) and 6.3 at 95\% CL \cite{HDBS-2022-03}.
Assuming \(\kthree=1\) then \(\kfour\) is restricted to be between \(-230\) and 240 at 95\% CL (both for expected and observed limits).


\FloatBarrier

\section{Conclusion}
\label{sec:conclusion}

This paper presents a search for triple-Higgs production in the \(b\bar{b}b\bar{b}b\bar{b}\) final state.
The search uses 126~\ifb of \(\pp\) collision data at \(\sqrt{s}=\SI{13}{\TeV}\) collected with the ATLAS detector at the LHC.
The data are interpreted using three different DNNs, which are each optimized to search for non-resonant, resonant, and heavy resonant (outside the perturbative unitarity bounds of the TRSM) signals with additional scalars.
The non-resonant interpretation includes a search for SM-like signals with coupling modifiers on the tri-linear and quartic Higgs self couplings \kthree~and \kfour.
The SM background is modeled using a data-driven extrapolation method, derived from the observed DNN score spectra in events with four or five \bjets to estimate the background shape in events with at least six \bjets.
The sensitivity of the search is impacted mainly by the statistical precision and the uncertainty in the data-driven extrapolation method.

No significant excess above the SM expectation is observed in the search for SM \(HHH\) production, nor is any significant excess observed in the search for various BSM signals with two additional heavy scalars \(X\) and \(S\).
A 95\% CL upper limit of 59 fb is set on the cross-section for SM \(HHH\) production.
The BSM searches include non-resonant production where \(\mS<250\) GeV, and resonant production up to \((\mX,\mS)=(1500,1000)\) GeV.


%
%
%
%
%
\section*{Acknowledgments}
%

%

%
%

%
%

We thank CERN for the very successful operation of the LHC and its injectors, as well as the support staff at
CERN and at our institutions worldwide without whom ATLAS could not be operated efficiently.

The crucial computing support from all WLCG partners is acknowledged gratefully, in particular from CERN, the ATLAS Tier-1 facilities at TRIUMF/SFU (Canada), NDGF (Denmark, Norway, Sweden), CC-IN2P3 (France), KIT/GridKA (Germany), INFN-CNAF (Italy), NL-T1 (Netherlands), PIC (Spain), RAL (UK) and BNL (USA), the Tier-2 facilities worldwide and large non-WLCG resource providers. Major contributors of computing resources are listed in Ref.~\cite{ATL-SOFT-PUB-2023-001}.

We gratefully acknowledge the support of ANPCyT, Argentina; YerPhI, Armenia; ARC, Australia; BMWFW and FWF, Austria; ANAS, Azerbaijan; CNPq and FAPESP, Brazil; NSERC, NRC and CFI, Canada; CERN; ANID, Chile; CAS, MOST and NSFC, China; Minciencias, Colombia; MEYS CR, Czech Republic; DNRF and DNSRC, Denmark; IN2P3-CNRS and CEA-DRF/IRFU, France; SRNSFG, Georgia; BMBF, HGF and MPG, Germany; GSRI, Greece; RGC and Hong Kong SAR, China; ICHEP and Academy of Sciences and Humanities, Israel; INFN, Italy; MEXT and JSPS, Japan; CNRST, Morocco; NWO, Netherlands; RCN, Norway; MNiSW, Poland; FCT, Portugal; MNE/IFA, Romania; MSTDI, Serbia; MSSR, Slovakia; ARIS and MVZI, Slovenia; DSI/NRF, South Africa; MICIU/AEI, Spain; SRC and Wallenberg Foundation, Sweden; SERI, SNSF and Cantons of Bern and Geneva, Switzerland; NSTC, Taipei; TENMAK, T\"urkiye; STFC/UKRI, United Kingdom; DOE and NSF, United States of America.

Individual groups and members have received support from BCKDF, CANARIE, CRC and DRAC, Canada; CERN-CZ, FORTE and PRIMUS, Czech Republic; COST, ERC, ERDF, Horizon 2020, ICSC-NextGenerationEU and Marie Sk{\l}odowska-Curie Actions, European Union; Investissements d'Avenir Labex, Investissements d'Avenir Idex and ANR, France; DFG and AvH Foundation, Germany; Herakleitos, Thales and Aristeia programmes co-financed by EU-ESF and the Greek NSRF, Greece; BSF-NSF and MINERVA, Israel; NCN and NAWA, Poland; La Caixa Banking Foundation, CERCA Programme Generalitat de Catalunya and PROMETEO and GenT Programmes Generalitat Valenciana, Spain; G\"{o}ran Gustafssons Stiftelse, Sweden; The Royal Society and Leverhulme Trust, United Kingdom.

In addition, individual members wish to acknowledge support from Armenia: Yerevan Physics Institute (FAPERJ); CERN: European Organization for Nuclear Research (CERN DOCT); Chile: Agencia Nacional de Investigaci\'on y Desarrollo (FONDECYT 1230812, FONDECYT 1230987, FONDECYT 1240864); China: Chinese Ministry of Science and Technology (MOST-2023YFA1605700, MOST-2023YFA1609300), National Natural Science Foundation of China (NSFC - 12175119, NSFC 12275265, NSFC-12075060); Czech Republic: Czech Science Foundation (GACR - 24-11373S), Ministry of Education Youth and Sports (FORTE CZ.02.01.01/00/22\_008/0004632), PRIMUS Research Programme (PRIMUS/21/SCI/017); EU: H2020 European Research Council (ERC - 101002463); European Union: European Research Council (ERC - 948254, ERC 101089007, ERC, BARD, 101116429), European Union, Future Artificial Intelligence Research (FAIR-NextGenerationEU PE00000013), Italian Center for High Performance Computing, Big Data and Quantum Computing (ICSC, NextGenerationEU); France: Agence Nationale de la Recherche (ANR-20-CE31-0013, ANR-21-CE31-0013, ANR-21-CE31-0022, ANR-22-EDIR-0002); Germany: Baden-Württemberg Stiftung (BW Stiftung-Postdoc Eliteprogramme), Deutsche Forschungsgemeinschaft (DFG - 469666862, DFG - CR 312/5-2); Italy: Istituto Nazionale di Fisica Nucleare (ICSC, NextGenerationEU), Ministero dell'Università e della Ricerca (PRIN - 20223N7F8K - PNRR M4.C2.1.1); Japan: Japan Society for the Promotion of Science (JSPS KAKENHI JP22H01227, JSPS KAKENHI JP22H04944, JSPS KAKENHI JP22KK0227, JSPS KAKENHI JP23KK0245); Norway: Research Council of Norway (RCN-314472); Poland: Ministry of Science and Higher Education (IDUB AGH, POB8, D4 no 9722), Polish National Agency for Academic Exchange (PPN/PPO/2020/1/00002/U/00001), Polish National Science Centre (NCN 2021/42/E/ST2/00350, NCN OPUS 2023/51/B/ST2/02507, NCN OPUS nr 2022/47/B/ST2/03059, NCN UMO-2019/34/E/ST2/00393, NCN \& H2020 MSCA 945339, UMO-2020/37/B/ST2/01043, UMO-2021/40/C/ST2/00187, UMO-2022/47/O/ST2/00148, UMO-2023/49/B/ST2/04085, UMO-2023/51/B/ST2/00920); Spain: Generalitat Valenciana (Artemisa, FEDER, IDIFEDER/2018/048), Ministry of Science and Innovation (MCIN \& NextGenEU PCI2022-135018-2, MICIN \& FEDER PID2021-125273NB, RYC2019-028510-I, RYC2020-030254-I, RYC2021-031273-I, RYC2022-038164-I); Sweden: Carl Trygger Foundation (Carl Trygger Foundation CTS 22:2312), Swedish Research Council (Swedish Research Council 2023-04654, VR 2018-00482, VR 2022-03845, VR 2022-04683, VR 2023-03403, VR grant 2021-03651), Knut and Alice Wallenberg Foundation (KAW 2018.0458, KAW 2019.0447, KAW 2022.0358); Switzerland: Swiss National Science Foundation (SNSF - PCEFP2\_194658); United Kingdom: Leverhulme Trust (Leverhulme Trust RPG-2020-004), Royal Society (NIF-R1-231091); United States of America: U.S. Department of Energy (ECA DE-AC02-76SF00515), Neubauer Family Foundation.

%
%


%
%
%
%
%
%

%
%
%
\clearpage
\printbibliography
\clearpage
\input{atlas_authlist}
%

%
%

%

%
%
%
%

\end{document}

%% file: atlas_authlist.tex
 
\begin{flushleft}
\hypersetup{urlcolor=black}
{\Large The ATLAS Collaboration}

\bigskip

\AtlasOrcid[0000-0002-6665-4934]{G.~Aad}$^\textrm{\scriptsize 104}$,
\AtlasOrcid[0000-0001-7616-1554]{E.~Aakvaag}$^\textrm{\scriptsize 17}$,
\AtlasOrcid[0000-0002-5888-2734]{B.~Abbott}$^\textrm{\scriptsize 123}$,
\AtlasOrcid[0000-0002-0287-5869]{S.~Abdelhameed}$^\textrm{\scriptsize 119a}$,
\AtlasOrcid[0000-0002-1002-1652]{K.~Abeling}$^\textrm{\scriptsize 56}$,
\AtlasOrcid[0000-0001-5763-2760]{N.J.~Abicht}$^\textrm{\scriptsize 50}$,
\AtlasOrcid[0000-0002-8496-9294]{S.H.~Abidi}$^\textrm{\scriptsize 30}$,
\AtlasOrcid[0009-0003-6578-220X]{M.~Aboelela}$^\textrm{\scriptsize 46}$,
\AtlasOrcid[0000-0002-9987-2292]{A.~Aboulhorma}$^\textrm{\scriptsize 36e}$,
\AtlasOrcid[0000-0001-5329-6640]{H.~Abramowicz}$^\textrm{\scriptsize 155}$
% ...
\bigskip
\\

% ...
$^{17}$Department for Physics and Technology, University of Bergen, Bergen; Norway.\\
% ...
$^{30}$Physics Department, Brookhaven National Laboratory, Upton NY; United States of America.\\
% ...
$^{36}$$^{(a)}$Facult\'e des Sciences Ain Chock, Universit\'e Hassan II de Casablanca;$^{(b)}$Facult\'{e} des Sciences, Universit\'{e} Ibn-Tofail, K\'{e}nitra;$^{(c)}$Facult\'e des Sciences Semlalia, Universit\'e Cadi Ayyad, LPHEA-Marrakech;$^{(d)}$LPMR, Facult\'e des Sciences, Universit\'e Mohamed Premier, Oujda;$^{(e)}$Facult\'e des sciences, Universit\'e Mohammed V, Rabat;$^{(f)}$Institute of Applied Physics, Mohammed VI Polytechnic University, Ben Guerir; Morocco.\\
% ...
$^{46}$Physics Department, Southern Methodist University, Dallas TX; United States of America.\\
% ...
$^{50}$Fakult\"{a}t Physik , Technische Universit{\"a}t Dortmund, Dortmund; Germany.\\
% ...
$^{56}$II. Physikalisches Institut, Georg-August-Universit\"{a}t G\"ottingen, G\"ottingen; Germany.\\
% ...
$^{104}$CPPM, Aix-Marseille Universit\'e, CNRS/IN2P3, Marseille; France.\\
% ...
$^{119}$$^{(a)}$New York University Abu Dhabi, Abu Dhabi;$^{(b)}$United Arab Emirates University, Al Ain; United Arab Emirates.\\
% ...
$^{123}$Homer L. Dodge Department of Physics and Astronomy, University of Oklahoma, Norman OK; United States of America.\\
% ...
$^{155}$Raymond and Beverly Sackler School of Physics and Astronomy, Tel Aviv University, Tel Aviv; Israel.\\
% ...
$^{a}$ Also Affiliated with an institute covered by a cooperation agreement with CERN.\\
$^{b}$ Also at An-Najah National University, Nablus; Palestine.\\
$^{c}$ Also at Borough of Manhattan Community College, City University of New York, New York NY; United States of America.\\
$^{d}$ Also at Center for High Energy Physics, Peking University; China.\\
$^{e}$ Also at Center for Interdisciplinary Research and Innovation (CIRI-AUTH), Thessaloniki; Greece.\\
$^{f}$ Also at CERN, Geneva; Switzerland.\\
% ...

\end{flushleft}
